\pdfoutput=1
\documentclass[apj]{emulateapj}
\usepackage[colorlinks,linkcolor=blue]{hyperref}
\usepackage{bm}
\usepackage{graphicx}
\usepackage{epsf}
\usepackage{graphics}
\usepackage{amsmath}
\usepackage{enumitem} 
\usepackage{multirow}
\usepackage{booktabs}
\usepackage{threeparttable}

\def\beq{\begin{equation}}
\def\eeq{\end{equation}}
\def\bey{\begin{eqnarray}}
\def\eey{\end{eqnarray}}

\def\Mpc{\,{\rm Mpc}}
\def\mpc{\, h^{-1}{\rm {Mpc}}}

\def\Msun{{\rm M_\odot}}
\def\msun{\, h^{-1}{\rm M_\odot}}

\def\gs{\mathrel{\raise1.16pt\hbox{$>$}\kern-7.0pt
\lower3.06pt\hbox{{$\scriptstyle \sim$}}}}
\def\ls{\mathrel{\raise1.16pt\hbox{$<$}\kern-7.0pt
\lower3.06pt\hbox{{$\scriptstyle \sim$}}}}
\def\gtsima{\, {\buildrel > \over \sim} \,}
\def\ltsima{\, {\buildrel < \over \sim} \,}
\def\prosima{\, {\buildrel \propto \over \sim} \,}
\def\gsim{\lower.5ex\hbox{\gtsima}}
\def\lsim{\lower.5ex\hbox{\ltsima}}
\def\simgt{\lower.5ex\hbox{\gtsima}}
\def\simlt{\lower.5ex\hbox{\ltsima}}
\def\simpr{\lower.5ex\hbox{\prosima}}

\def\Mhalo{M_{\rm halo}}

\shorttitle{ELUCID VI: Cosmic variance of galaxy distribution in the local Universe}
\shortauthors{Chen Yangyao et al.}

\begin{document}

\title {ELUCID. VI: Cosmic variance of galaxy distribution in the local Universe}
\author{
Yangyao Chen\altaffilmark{1},
H.J. Mo \altaffilmark{1,2}, 
Cheng Li\altaffilmark{1},
Huiyuan Wang\altaffilmark{3,4}, 
Xiaohu Yang\altaffilmark{5,6},
Shuang Zhou\altaffilmark{1},
Youcai Zhang\altaffilmark{7}
}
\altaffiltext{1}{Center for Astrophysics and Physics Department, Tsinghua University, Beijing 100084, China; yangyao-17@mails.tsinghua.edu.cn}
\altaffiltext{2}{Department of Astronomy, University of Massachusetts, Amherst MA 01003-9305, USA}
\altaffiltext{3}{Key Laboratory for Research in Galaxies and Cosmology, Department of Astronomy, University of Science and
Technology of China, Hefei, Anhui 230026, China}
\altaffiltext{4}{School of Astronomy and Space Science, University of Science and Technology of China, Hefei 230026, China}
\altaffiltext{5}{Department of Astronomy, Shanghai Jiao Tong University, Shanghai 200240, China}
\altaffiltext{6}{IFSA Collaborative Innovation Center, Shanghai Jiao Tong University, Shanghai 200240, China}
\altaffiltext{7}{Shanghai Astronomical Observatory,
Shanghai 200030, China}

\begin{abstract}
Halo merger trees are constructed from ELUCID, a constrained $N$-body 
simulation in the Sloan Digital Sky Survey (SDSS) volume. These merger
trees are used to populate dark matter halos with galaxies according  
to an empirical model of galaxy formation. Mock catalogs in the SDSS 
sky coverage are constructed, which can be used to study the spatial 
distribution of galaxies 
in the low-$z$ Universe. These mock catalogs are used to quantify     
the cosmic variance in the galaxy stellar mass function (GSMF) measured 
from the SDSS survey. The GSMF estimated from the SDSS 
magnitude-limited sample can be affected significantly by the 
presence of the under-dense region at $z<0.03$, so that the low-mass 
end of the function can be underestimated significantly. Several 
existing methods designed to deal with the effects of 
the cosmic variance in the estimate of GSMF are tested, and none 
is found to be able to fully account for the cosmic variance. 
We propose a method based on the conditional stellar mass functions  
in dark matter halos, which can provide an unbiased 
estimate of the global GSMF. The application of the method to 
the SDSS data shows that the GSMF has a significant upturn at 
$M_*< 10^{9.5} \msun$, which has been missed in many
earlier measurements of the local GSMF.
\end{abstract}


%%%%%%%%%%%%%%%%%%%%%%%%%%%%%%%%%%%%%%%%%%%%%%
%%%%%%%%%%%%% Part I. Intro %%%%%%%%%%%%%%%%%%
%%%%%%%%%%%%%%%%%%%%%%%%%%%%%%%%%%%%%%%%%%%%%%
% Yangyao, add in 1020/2017
% Yangyao, update, 0807/2018
% HJM revised  8/11/2018
\section{Introduction}
\label{sec_intro}
The Universe contains prominent structures up to $\sim 100\Mpc$, only 
reaching homogeneity on much large scales \citep[e.g.][]{Peebles1980,Davis1985}.
The properties of galaxies and other objects, which form and evolve in 
the cosmic web, are expected to be affected by their large-scale 
environments. Thus, astronomical observations, which are always made 
in limited volumes in the Universe, can be affected by the cosmic variance
(CV) caused by spatial variations of the statistical properties  
of cosmic objects, such as galaxies, due to the presence 
of large scale structure. Because of CV, statistics obtained 
from a sample that covers a specific volume in the Universe may be 
different from those expected for the Universe as a whole. 
Erroneous inferences would then be made if such biased observational 
data were used to constrain models. 

Cosmic variance (CV) is a well known problem 
\citep[e.g.][]{Somerville2004,Jha2007,Driver2010,Moster2011,Marra2013,Keenan2013,Wojtak2014,Whitbourn2014,Whitbourn2016}, and various 
attempts have been made to deal with it. 
One way is to analyze different (sub-)samples, e.g. obtained from the Jackknife sampling of 
the observational data, and to use the variations among them to have some handle on 
the CV. However, this can only provide information about the variance 
within the total sample itself, but not that of the total sample relative to a fair 
sample of the Universe. 
Another way is to use the spatial distribution of bright galaxies
(a density-defining population), which can be observed in a large volume, to
quantify the CV expected in sub-volumes \citep[e.g.][]{Driver2010}, or to 
re-scale (or correct) the number density of faint galaxies observed in a smaller 
volume, as was done by \citet{Baldry2012} in their estimate of galaxy stellar mass 
functions in the GAMA sample. However, this method relies on the assumption 
that galaxies of different luminosities/masses have similar spatial distributions, 
which may not be true. The same problem also exists in the maximal likelihood method 
\citep[e.g.][]{G.Efstathiou1988}, where galaxy luminosity function is explicitly 
assumed to be independent of environment. Yet another way is to estimate
the CV expected from a given sample using simple, analytic models 
for the clustering properties of galaxies on large scales. 
Along this line, \citet{Somerville2004} tested the 
effects of the CV on different scales, and proposed the use of either 
the two-point correlation function of galaxies, or the combination of the 
linear density field with halo bias models \citep[e.g.][]{MoWhite1996,Sheth2001},
to predict the CV of different surveys. Similarly, \citet{Moster2011} carried 
out an investigation of the CV expected in observations of the galaxy populations 
at different redshifts, using the linear density field predicted by the 
${\rm \Lambda CDM}$ model combined with a bias model that takes into 
account the dependence of galaxy distribution on galaxy mass and redshift. 
Unfortunately, such an approach does not take into account observational 
selection effects.  More importantly, this approach only gives a 
statistical estimate of the CV but does not measure the deviation of a specific
sample from a fair sample. Finally, one can also use a large number 
of mock galaxy samples, either obtained directly from hydrodynamic 
simulations, or from N-body simulation-based semi-analytic 
(SAM) and empirical models, to quantify how the sample-to-sample
variation of the statistical measure in question depends on sample volume. 
However, this needs a large set of simulations for each model, analyzed 
in a way that takes into account the observational selection effects 
in the data, which in practice is costly and time consuming. 
Furthermore, the same as the approach based on galaxy clustering statistics, 
this approach can only provide a statistical statement of the 
expected CV, but does not provide a way to correct the variance 
of a specific sample. 

Can one develop a systematic method to study the cosmic 
variance, and to quantify and correct biases that are present in 
observational data? The answer is yes, and the key is to use constrained 
simulations. Indeed, if one can accurately reconstruct the initial conditions 
for the formation of the structures in which the observed galaxy population reside, 
one can then carry out simulations with such initial conditions in a sufficiently
large box that contains the constrained volume, so that the large box can be used 
as a fair sample, while the constrained region can be used to model the 
observational data. By comparing the statistics obtained from the mock samples with 
those obtained from the whole box, one can quantify and correct the 
CV in the observational data. 

In the past few years, the ELUCID collaboration has embarked on 
the development of a method to accurately reconstruct the initial conditions 
responsible for the density field in the low-$z$ Universe
\citep{Wang2014}. As demonstrated by various tests
\citep{Wang2014,Wang2016}, the reconstruction method is much more accurate 
than other methods that have been developed, and works reliably even 
in highly non-linear regimes. The initial conditions in a $500\mpc$ box
that contains the main part of the SDSS volume have already been obtained, 
and a high resolution $N$-body simulation, run with $(3072)^3$ particles, 
has been carried out with these initial conditions in the current 
$\Lambda$CDM cosmology \citep{Wang2016}.

In the present paper, we use the dark matter halo merger trees constructed 
from the ELUCID simulation to populate simulated halos with model galaxies 
predicted by the empirical galaxy formation model developed 
by~\citet[][thereafter L14, L15]{Lu2014,Lu2015}. 
The model galaxies in the constrained volume are then used to construct mock 
catalogs that contain the same CV as the real SDSS sample. 
We compare galaxy stellar mass functions (GSMF) estimated from the mock catalogs 
with that obtained from the total simulation box to quantify the CV within the
SDSS volume. Finally, we propose a method based on the conditional stellar mass 
or luminosity distribution in dark matter halos to correct for the CV in the 
observed GSMF. As we will see, the CV can be very severe in the low-mass end of 
the GSMF obtained from methods commonly adopted, and the low-mass end slope of
the true GSMF in the low-$z$ Universe may be significantly steeper than 
those published in the literature. 

The structure of the paper is as follows. 
In \S\ref{sec_mergingtrees}
we describe methods to implement Monte Carlo halo merger trees 
in simulated merger trees, so as to extend all trees 
down to a mass resolution sufficient for our purpose. 
In \S\ref{sec_populating} we populate simulated halos with galaxies 
using an empirical model of galaxy formation,  
and construct a number of mock catalogs to mimics the SDSS
survey both in spatial distribution and physical properties. 
In \S\ref{sec_CVinGSMF} we examine in detail the cosmic variance 
in the estimates of the GSMF, and show how commonly adopted 
methods to measure the galaxy luminosity function (GLF) and 
GSMF fail to account for the CV. We also propose and test 
a new method to correct for CV in GLF and GSMF, and apply 
it to the real SDSS data to obtain the CV-corrected GLF and 
GSMF. Finally, a brief summary of our main results is presented in 
\S\ref{sec_summary}.

Throughout the paper, we define the GSMF as 
$\Phi(M_*)=\mathrm{d}N/\mathrm{d}V/\mathrm{d}\log M_*$, which is 
the number of galaxies per unit volume per unit stellar mass in 
logarithmic space, and define the GLF in $\rm X$-band as 
$\Phi(M_{\rm X})={\rm d} N/{\rm d}V/{\rm d}(M_{\rm X}-5\log h)$, 
which is the number of galaxies per unit volume per unit 
magnitude. The magnitude $M_{\rm X}$ is $k$-corrected to redshift 
$0.1$ without evolution correction, unless specified otherwise.

%%%%%%%%%%%%%%%%%%%%%%%%%%%%%%%%%%%%%%%%%%%%%%
%%%%%%%%%%%%% Part II. Halo %%%%%%%%%%%%%%%%%%
%%%%%%%%%%%%%%%%%%%%%%%%%%%%%%%%%%%%%%%%%%%%%%
% HJM has revised this section

\section{Merger trees of dark matter halos from the ELUCID simulation}
\label{sec_mergingtrees}

\subsection{The simulation}

We use the ELUCID simulation carried out by \citet{Wang2016} to model 
the dark matter halo population, their formation histories, and 
spatial distribution. This is an $N$-body simulation that uses L-GADGET, 
a memory optimized version of GADGET-2~\citep{Springel2005}, to 
follow the evolution of $3072^3$ dark matter particles 
(each with a mass of $3.088 \times 10^8 \msun$) in a periodic cubic box with 
side length of $500\mpc$ in co-moving units. 
The cosmology used is the one based on WMAP5~\citep{Dunkley2009,Komatsu2009}: 
a flat Universe with $\Omega_K=0$;
a matter density parameter $\Omega_{\rm m,0}=0.258$; 
a cosmological constant $\Omega_{\Lambda,0}=0.742$;
a baryon density parameter $\Omega_{\rm B,0}=0.044$; 
a Hubble constant $H_0=100h\ \mathrm{km\ s^{-1}\ \Mpc^{-1}}$ with $h=0.72$; and a Gaussian initial density field with power spectrum $P(k)\propto k^n$,
with $n=0.96$ and with the amplitude specified by $\sigma_8=0.80$. 
The simulation is run from redshift $z=100$ to $z=0$, 
with outputs recorded at 100 snapshots between $z=18.4$ and $z=0.0$. 

The initial conditions (phases of Fourier modes) of the density field are 
those obtained from the reconstruction based on the halo-domain method 
of~\citet{Wang2009} and the Hamiltonian Markov Chain Monte Carlo
(HMC) method~\citep{Wang2013,Wang2014}, 
constrained by the distributions of dark matter halos represented by galaxy 
groups and clusters selected from the SDSS redshift survey \citep{Yang2007,Yang2012}. 
As shown in \citet{Wang2016} with the use of mock catalogs, more than $95\%$ of 
the groups with masses above $10^{14}\msun$ can be matched with the simulated 
halos of similar masses, with a distance error tolerance of $\sim 4\mpc$, and 
massive structures such as the Coma cluster and the Sloan Great Wall
can be well reproduced in the reconstruction. Thus, the use of the constrained 
simulation from ELUCID allows us not only to model accurately the large-scale 
environments within which observed galaxies reside, but also to
recover, at least partially, the formation histories of the massive 
structures seen in the local Universe.     

\subsection{The construction of halo merger trees}
\label{ssec_trees}

\begin{figure*}
\centering
 \includegraphics[width=12cm]{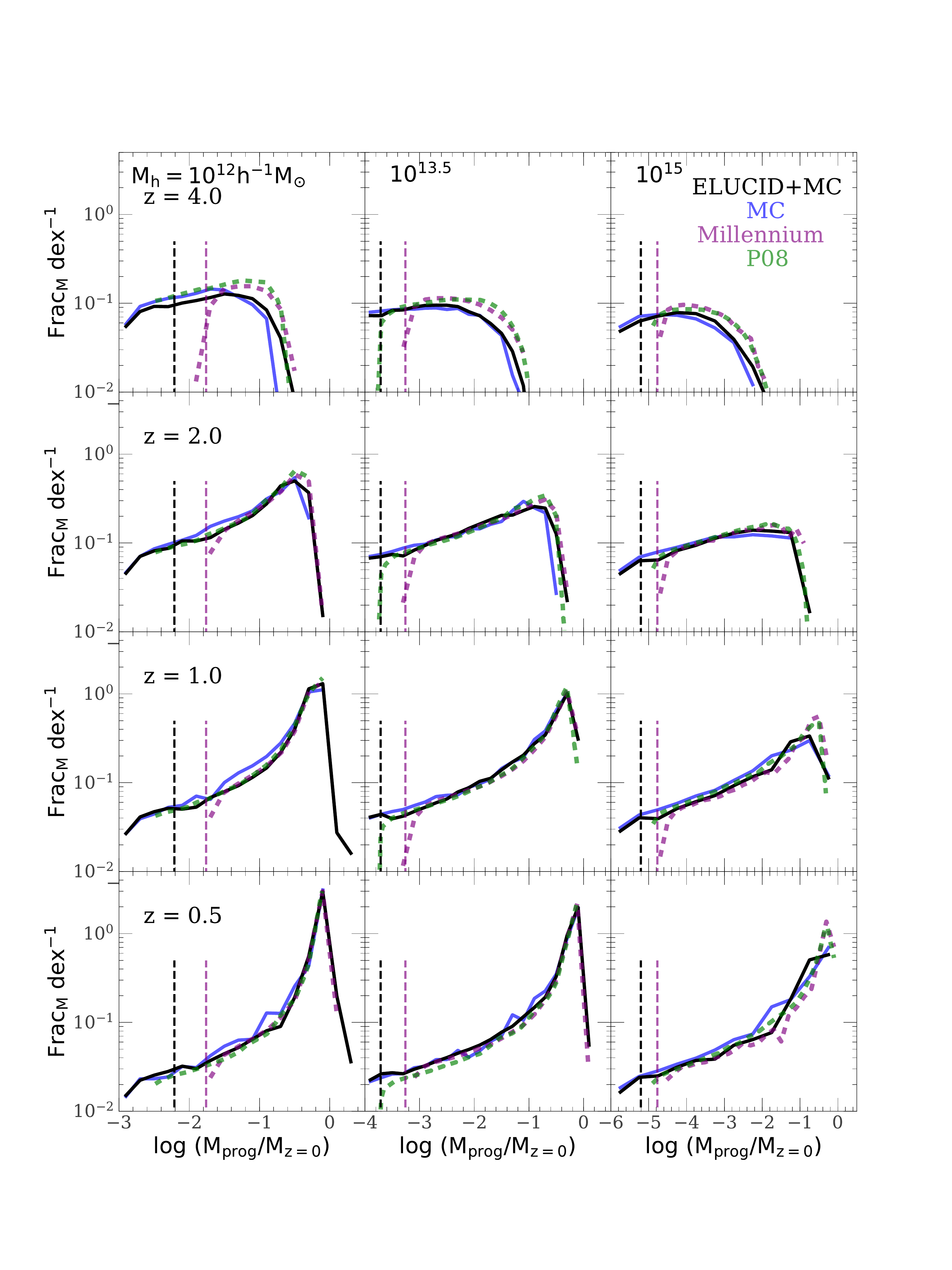}
 \caption{
 Conditional progenitor mass functions (mean fraction of mass in progenitors in per unit 
 progenitor mass $M_{\rm prog}/M_{z=0}$ bin in logarithmic space) of dark matter halos from 
 different kinds of merging trees, for different $z=0$ halos of different masses 
 (each column) and for progenitors at different redshifts (each row).
 Black solid: halo merger trees obtained from ELUCID simulation 
 repaired by Monte-Carlo-based trees. 
 Blue solid: P08~\citep{Parkinson2007} Monte Carlo trees generated with the WMAP5 cosmology. 
 Purple dashed: Millennium~\citep{Springel2005} FOF halo merger trees.
 Green dashed: P08 Monte Carlo trees with Millennium cosmology.
 The two vertical solid lines show the 20 particles mass resolution of halos in 
 ELUCID and Millennium simulations, respectively.}
 \label{fig:CHMF}
\end{figure*}

Halos and their sub-halos with more than 20 particles are identified with the 
friend-of-friend (FOF) and SUBFIND algorithms \citep{Springel2005b}.
To be safe, we only use halos identified in the simulation with masses 
$M_{\rm h} \geq M_{\rm th} = 10^{10}\msun$. 
However, this mass resolution is not 
sufficient to resolve lower mass halos in which star formation may still be 
significant, particularly at high $z$. In order to trace the star formation 
histories in halos to high redshifts, we need to reach a halo mass of 
about $10^9\msun$, below which star formation is expected to be unimportant due to photo-ionization heating 
\citep[e.g.][]{Babul1992,Thoul1996}. Here we adopt a Monte Carlo method to extend 
the merger trees of the simulated halos down to a mass limit, 
$10^{9}\msun$. \citet{Jiang2014} have tested the performances of 
several different methods of generating Monte Carlo halo merger trees, 
and found that the method of \citet[][thereafter P08]{Parkinson2007}
consistently provides the best match to the halo merging trees obtained from 
$N$-body simulations. We therefore adopt the P08 method.  

We join the P08 Monte Carlo trees to the halo merger trees obtained 
from the simulation through the following steps: 
\begin{enumerate}[fullwidth,itemindent=1em,label=(\roman*)]
\item 
For each simulated halo merger tree $T$, we eliminate halos that have masses 
below $M_{\rm th} = 10^{10} \msun$ but have no progenitors more massive than 
$M_{\rm th}$. The purpose of the second condition is to preserve halos which once 
had masses larger than $M_{\rm th}$ but have become less massive later 
due to stripping and/or mass loss. 
\item 
For each halo $H$ that is not eliminated in $T$, we generate a Monte Carlo 
tree $t$ (down to $10^9\msun$), rooted from a halo $h$ that has the 
same mass and the same redshift as $H$, and eliminate all halos more 
massive than $10^{10}\msun$ in $t$.
\item 
We add $t$ to $H$. The procedure is repeated for all halos with masses above 
$10^{10}\msun$ in all trees in the ELUCID simulation, 
so that all such halos have merger trees extended to $10^9\msun$. 
\item
For halos with masses below $10^{10}\msun$ at $z=0$, their merger trees are
entirely generated with the Monte Carlo method. Note that these 
halos are not identified from the simulation, but can be used 
to model galaxies in such low-mass halos when needed.
\end{enumerate}

With these steps, we obtain `repaired' halo merger trees that have 
a mass resolution of $10^9\msun$, with halos more massive than 
$10^{10}\msun$ sampled entirely by the simulation, and 
the less massive ones modeled by Monte-Carlo trees.
Fig.~\ref{fig:CHMF} shows the conditional progenitor mass functions of 
dark matter halos, defined as the fraction of mass in progenitors per 
logarithmic mass, for merger trees rooted from different masses, and for 
progenitors at different redshifts. Our results, obtained by combining the 
simulated trees above the mass resolution $M_{\rm th}$ with the Monte Carlo 
merging trees generated with 
the P08 model below the mass limit, are shown by the black solid lines, and compared 
with the merger trees generated entirely with the P08 model.  
Overall, the progenitor mass distributions we obtain match well 
those obtained from the Monte Carlo method, 
indicating that our merger trees are reliable.

Since galaxies form and evolve in dark matter halos, our `repaired' halo merger 
trees from the ELUCID simulation provide the basis to link galaxy properties to 
dark matter halos, and can be used in combinations with halo-based methods 
of galaxy formation, such as abundance matching, semi-analytic and other 
empirical models, to populate halos with galaxies. The method can, 
in principle, be applied to simulated halos with any mass resolution and 
with any cosmology, to extend halo merger trees to a sufficiently low mass, 
as long as reliable Monte Carlo trees can be generated.
We note that our merger trees do not include high order 
sub-halos, i.e. sub-halos in sub-halos.
In the next section, we apply the empirical model, developed 
in L14 and L15, to follow galaxy formation and evolution in dark matter 
halos, based on our repaired halo merger trees.

%%%%%%%%%%%%%%%%%%%%%%%%%%%%%%%%%%%%%%%%%%%%%%%%%
%%%%% Part III. Populating halos with galaxies %%
%%%%%%%%%%%%%%%%%%%%%%%%%%%%%%%%%%%%%%%%%%%%%%%%%

\section{Populating halos with galaxies}
\label{sec_populating}

In this section, we describe the L14, L15 empirical method, developed by 
\citet{Lu2014,Lu2015}, to populate galaxies in the halo merger trees described 
in the previous section. Briefly, we assign a central galaxy to each 
distinctive halo and give it an appropriate star formation rate 
(SFR) according to the empirical model. We then evolve all galaxies
in the current snapshot to the next, following the accretion of 
galaxies by dark matter halos and the mergers of galaxies.
The stellar masses for both central and satellite galaxies are obtained 
by integrating the stellar contents along their histories.  Finally, 
observable quantities, such as luminosity and apparent magnitude, are 
obtained from a stellar population synthesis model.   

\subsection{The empirical model of galaxy formation}
\begin{figure}
\includegraphics[width=\columnwidth]{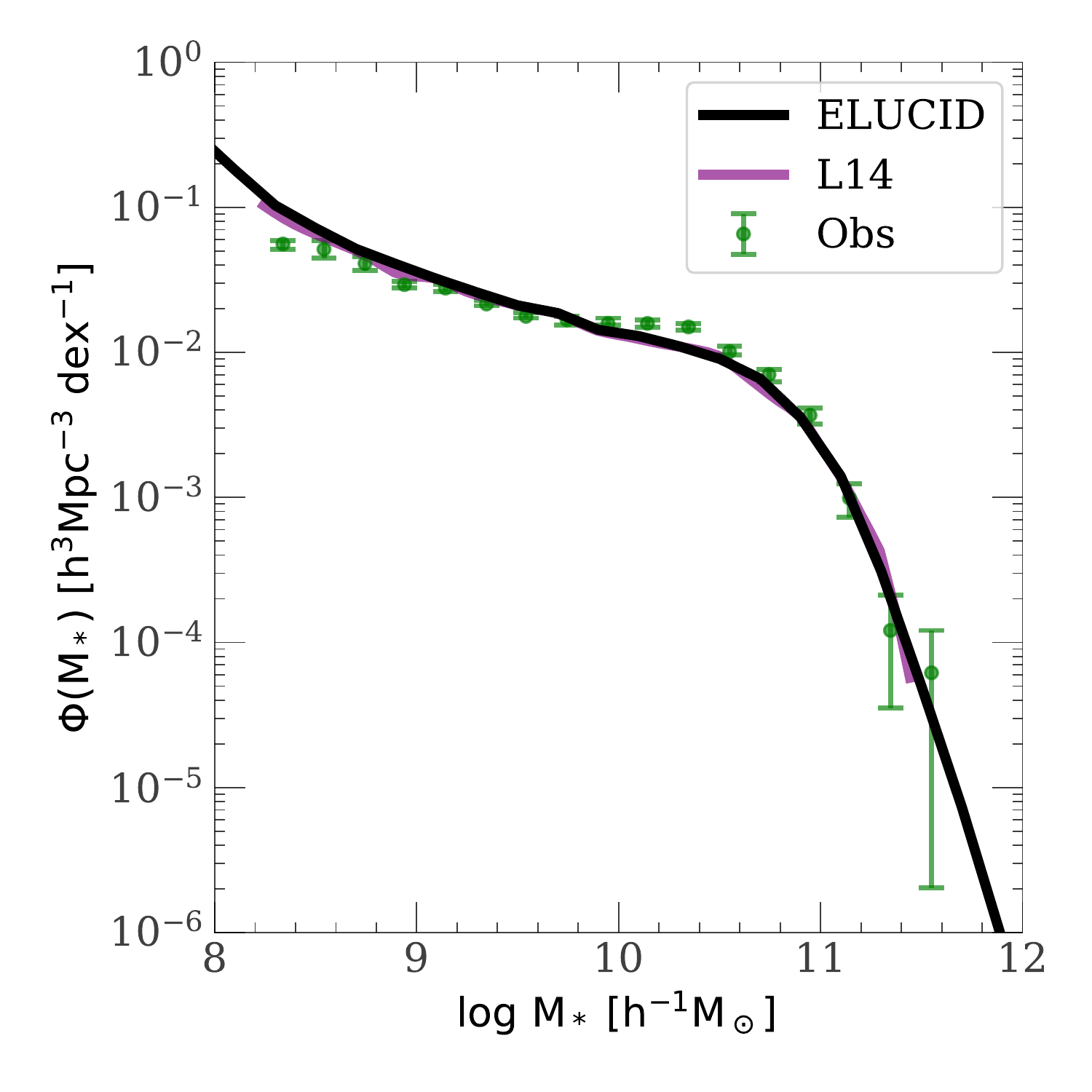}
\caption{ Galaxy stellar mass functions at redshift $z=0$. Black solid line: model galaxies based on our repaired trees. Purple solid line: from ~\citealp{Lu2014}, based on Monte Carlo halo merger trees. Green dots with error bars: from observational result which is used by L14 to calibrate the model.}
\label{fig:gsmf}
\end{figure}

\begin{figure}
 \includegraphics[width=\columnwidth]{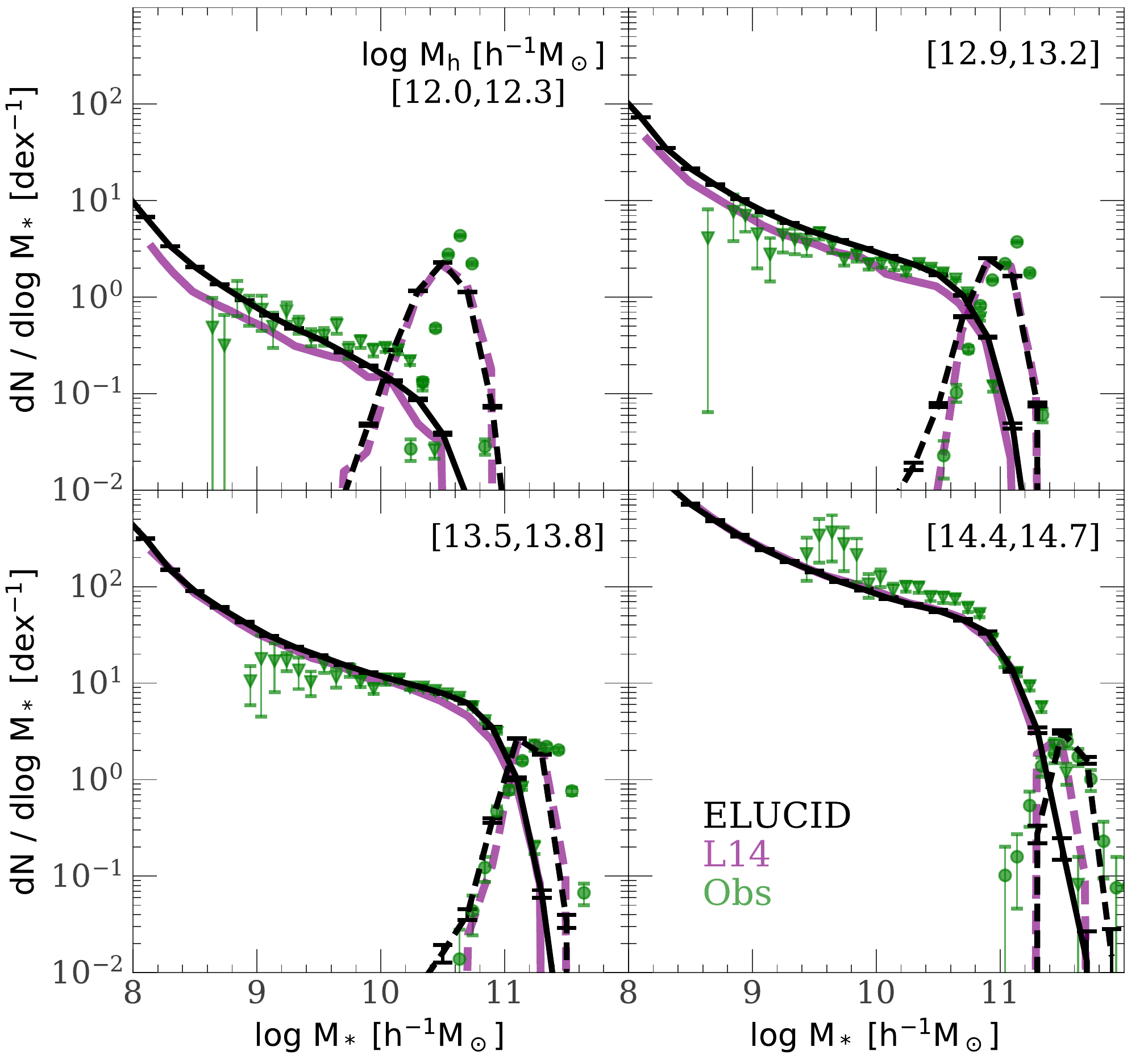}
 \caption{Conditional stellar mass functions in halos with different mass $M_{\rm h}$ (as indicated in each panel) at redshift $z=0$. Black lines: model galaxies based on our 
 repaired trees, for central galaxies (dashed) and satellite galaxies (solid). 
 The error bars indicate the standard deviations among 100 bootstrap resamplings.
 Purple lines: from~\citealp{Lu2014}, 
 based on Monte Carlo halo merger trees, for central galaxies (dashed) and 
 satellite galaxies (solid).
 Green markers with error bars: from observational result of~\citealp{Yang2008}, for central galaxies (circles) and satellite galaxies (triangles).
 }
 \label{fig:cgsmf}
\end{figure}

\begin{figure*}
\centering
 \includegraphics[width=15cm]{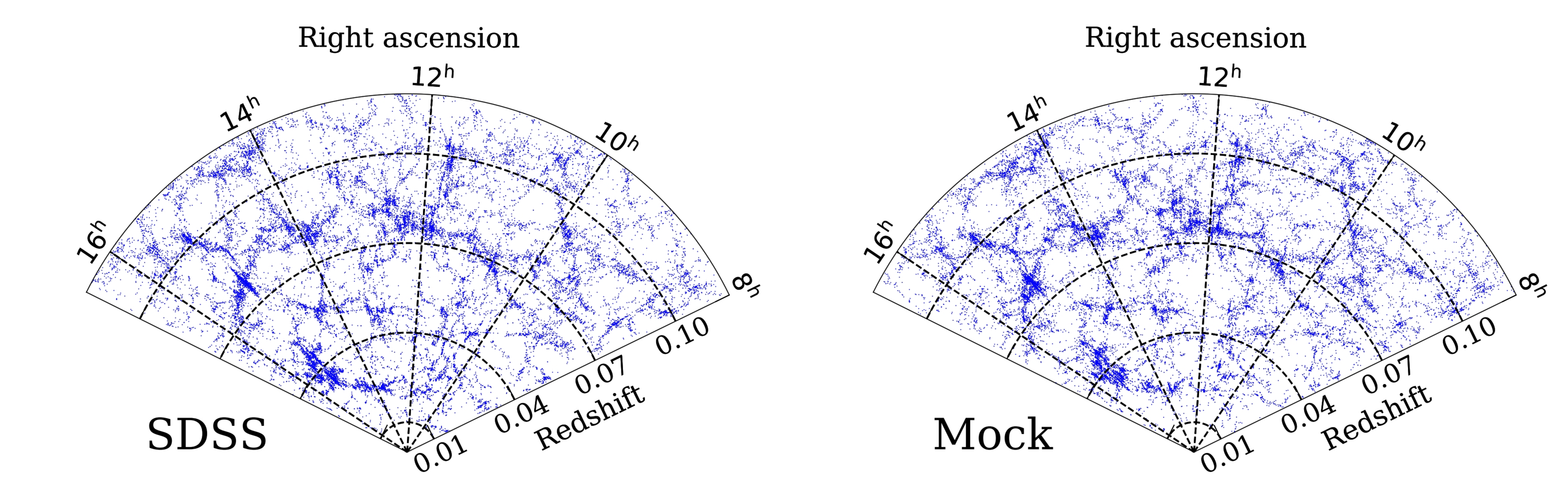}
 \caption{Spatial distribution of SDSS (left panel) and mock (right panel) galaxies. 
 Selections are made for all galaxies with $r\le 17.6$ in the 
 redshift range $[0.01,\ 0.12]$ in the Sloan NGC region. Only galaxies 
 in a $4^\circ$ declination slice are plotted.}
 \label{fig:sky_sdss}
\end{figure*}

In the model of L14 and L15, SFR of a central 
galaxy is assumed to depend on the halo mass $\Mhalo$ and redshift $z$ as
\beq
\begin{split}
{\rm SFR}&(\Mhalo,z) =\\
&\varepsilon \frac{f_{\rm B} \Mhalo}{\tau _0}
(1+z)^{\kappa}(1+X)^{\alpha}
\left(\frac{X+R}{X+1}\right)^{\beta}
\left(\frac{X}{X+R}\right)^{\gamma}
\end{split}
\eeq
where $\tau_0 = 1/(10 H_0)$, $\kappa = 3/2$. $f_{\rm B}=\Omega_{\rm B,0}/\Omega_{\rm m,0}$ is the cosmic 
baryon fraction, and $\varepsilon$ and $\beta$ are time-independent model 
parameters. The parameters, $\alpha$ and $\gamma$, are assumed to be 
time-dependent, given by  
\beq
\alpha = \alpha_0(1+z)^{\alpha^{'}} 
\eeq
and 
\beq
\gamma = \left\{
\begin{aligned}
&\gamma_a& \mathrm{if}\ z < z_c\\
&(\gamma_a-\gamma_b)(\frac{z+1}{z_c+1})^{\gamma^{'}} + 
\gamma_b & \mathrm{if}\ z \geq z_c
\end{aligned}
\right.
\eeq
where $\alpha_0$, $\alpha^{'}$, $\gamma_a$, $\gamma_b$ and $z_c$ are time-independent 
model parameters. In L14 and L15, both $\alpha$ and $\gamma$ 
are chosen to be time-dependent to make the model compatible with the observed 
galaxy stellar mass functions (GSMFs) at different redshifts and the composite 
conditional luminosity function of cluster galaxies at redshift $z=0$~\citep[see also][]{Lim2017}. 
All the model parameters are determined by fitting the model predictions to a 
set of observational data (see the original papers for details). Here we 
adopt the parameters listed in L14 (denoted by 'Model III SMF+CGLF' in this paper), 
which are based on a cosmology consistent
with the WMAP5 cosmology~\citep{Dunkley2009,Komatsu2009} used here. 

Once a dark matter halo hosting a galaxy is accreted by a bigger halo, 
the central galaxy in it is assumed to become a satellite galaxy, and thus 
experience satellite-specific processes, such as tidal stripping and ram-pressure 
stripping, which may reduce and quench its star formation. L14 modelled 
the SFR in satellites as 
\beq
{\rm SFR}(M_*,z) = {\rm SFR}(t_{\rm accr})
\exp\left[ - \frac{t-t_{\rm accr}} {\tau(M_*)}\right]
\eeq
with 
\beq
\tau(M_*) = \tau_{*,0}\exp(-M_*/M_{*,c})\,.
\eeq
Here $M_*$ is the current mass of the satellite galaxy,
$\tau_{*,0}$ and $M_{*,c}$ are time-independent model parameters, and $t_{\rm accr}$ is the cosmic 
time at which the host halo of the galaxy is accreted.  
After accretion, the satellite halo and the galaxies it hosts 
are expected to experience dynamical friction, which causes them to
move towards the inner part of the new host halo. The satellites 
may then merger with the central galaxy located near the center.   
We follow L14 and use an empirical model to determine the time 
when the merger occurs:
\beq
\Delta t = 0.216\frac{Y^{1.3}}{\ln(1+Y)}\exp(1.9\eta)\frac{r_{\rm halo}}{v_{\rm halo}},
\eeq
where $Y=M_{\rm cen}/M_{\rm sat}$ is the ratio of mass between 
the central halo and the satellite halo at the time 
when the accretion occurs, 
and $r_{\rm halo}$ and $v_{\rm halo}$ are the virial radius and 
virial velocity of the central halo \citep[e.g.][]{Boylan-Kolchin2008}.
The parameter, $\eta$, describes the specific orbital angular momentum, 
and is assumed to follow a probability distribution 
$P(\eta) = \eta^{1.2}(1-\eta)^{1.2}$ \citep[e.g.][]{Zentner2005}.
After merger, a fraction of $f_{\rm TS}$ of the stellar mass of the 
satellite is added to the central galaxy, with $f_{\rm TS}$ a model 
parameter. 

The ingredients given above can be used to predict the stellar mass 
and SFR of both central and satellite galaxies. In order to make predictions 
for galaxy luminosities in different bands, we also need the metallicities 
of stars. We use the mean metallicity - stellar mass relation given by 
\citet{Gallazzi2005} to assign metallicities to galaxies according to 
their masses. A simple stellar population synthesis model, based on 
the \cite{Bruzual2003} with a Chabrier initial mass 
function \citep{Chabrier2003}, 
is adopted to obtained  the mass to light ratio of formed star, and the 
mass loss due to stellar evolution.

We note that the L14 model, which is based on Monte-Carlo merger trees, 
does not take into account some special events that exist in numerical 
simulations. In simulated merger trees, some sub-halos 
were main halos at some early times, accreted into other systems 
as satellites later, and were eventually ejected and became main 
halos again. For such cases, we treat the galaxy in the sub-halo
as a satellite galaxy even after the sub-halo is ejected. The ejected 
sub-halos are then treated as new main halos after ejection. 
This implementation does not make much physical sense, but  
best mimic the Monte Carlo merger trees in which sub-halos are never 
ejected, and all halos at a given time are treated equally without 
depending on whether or not they have gone through a big halo. 
Such an implementation is necessary, as the model parameters given by 
L14 are calibrated by using Monte-Carlo merger trees.  

Fig.~\ref{fig:gsmf} shows the galaxy stellar mass function (GSMF) of  
model galaxies at redshift z = 0, in black solid line, in comparison with 
the result of L14 (purple solid line). As one can see, the L14 result is well 
reproduced over wide ranges of stellar masses, which 
demonstrates that our implementation of the L14 model with the ELUCID halo 
merger trees are reliable, as long as the general galaxy GSMF is concerned. For reference, we also include the 
observational data points (green dots with error bars) that were used in 
L14 to constrain their model parameters. 

As a more demanding test, we compare in Fig.\,\ref{fig:cgsmf}
the conditional galaxy stellar mass functions (CGSMFs) in halos of 
different masses at redshift $z=0$ obtained from the ELUCID halo merger trees with 
those given by L14. Here again we see a good agreement between the 
two. Since the CGSMF gives the average number of galaxies of a given 
stellar mass in a halo of a given mass, a good match in 
CGSMFs also implies that the spatial clustering of galaxies as 
a function of stellar mass is also reproduced.

\subsection{Galaxy occupation in dark matter halos}

To use our model galaxies to construct mock catalogs, we need to assign 
spatial positions and peculiar velocities to galaxies in each halo in the 
simulation according to the halo occupation distributions (HODs) 
obtained from the 
empirical model described above. Here we adopt a sub-halo abundance matching 
method that links galaxies in a halo to the sub-halos in it. As shown in 
\citet{Wang2016}, the sub-halo population can be identified reliably from 
the ELUCID simulation for sub-halos with masses 
down to $\sim 10^{10}\msun$. The abundance matching goes as follows.  
For a given halo, we first rank galaxies in descending order of stellar mass
and sub-halos in descending order of halo mass. Here the mass of a sub-halo 
is that at the time when the sub-halo was first accreted into its host. 
Note that sub-halos both identified directly from the simulation and 
added using Monte Carlo merger trees 
%\footnote{The galaxies hosted by Monte-Carlo sub-halos are usually too faint to be relevant; they are included for completeness.} 
(see~\S\ref{ssec_trees}) are used. For sub-halos identified in the simulation, 
their positions and velocities are those given by the SUNFIND.
For the Monte Carlo sub-halos that are joined to the simulated halos, 
on the other hand, we assign random positions according to the NFW 
profile \citep{Navarro1997} with concentration parameters given by 
\citet{Zhao2009}, and their velocities are drawn from a Gaussian distribution with 
dispersion appropriate for the density profile assumed. 
For sub-halos that are rooted from a $z=0$ Monte Carlo halo, the galaxies 
hosted by them are usually too faint to be relevant; they are only included 
for completeness, but actually are not used in constructing the mock catalog.
Finally, the position and velocity of the sub-halo are 
assigned to the galaxy that has the same rank. For those galaxies that do not have 
sub-halo counter-parts, their positions and velocities are assigned 
randomly according to the NFW profile.
This method can be used to construct volume limited 
samples within the entire simulation box down to stellar 
masses $\sim 10^8\msun$, with full phase space information 
obtained from the simulated sub-halos. This is sufficient for most 
of our purposes.  

\subsection{The SDSS mock catalog}

With full information about the luminosities and phase space coordinates 
for individual galaxies, it is straightforward to make mock catalogs using 
galaxies in the constrained volume and applying the same selection criteria as 
in the observation. For each model galaxy in the simulation box, we 
assign to it a cosmological redsfhit, $z_{\rm cos}$, according to  
its distance to a virtual observer, and the observed redshift, 
$z_{\rm obs}$, is given by $z_{\rm cos}$ together with its
line-of-sight (los) peculiar velocity, $v_{\rm los}$:
\beq
z_{\rm obs} = z_{\rm cos} + (1+z_{\rm cos})\frac{v_{\rm los}}{c}\,,
\eeq
with $c$ the speed of light. Here the location of the 
virtual observer and the coordinate system are determined 
by the orientation of the SDSS volume in the simulation box. 
SDSS apparent magnitudes in $u$, $g$, $r$, $i$, and $z$ are assigned to 
each galaxy according to its luminosities in the corresponding 
bands. For our SDSS mock sample,  we select all galaxies in 
the SDSS Northern-Galactic-Cap (NGC) region with redshifts 
$0.01 < z < 0.12$ and with magnitude $r \leq 17.6$. 

Fig.~\ref{fig:sky_sdss} shows the real SDSS galaxies (left) and 
the mock galaxies (right) in the same slice in the SDSS sky coverage. 
It is clear that the distribution of the mock galaxies is very 
similar to that of the real galaxies. The large scale structures in the 
local Universe, such as the Sloan Great Wall at redshift 
$\approx 0.08$, are well reproduced. Thus, the mock catalog can 
be used to investigate both the properties of the galaxy population 
in the cosmic web, and the large scale clustering  of galaxies. 
In particular, since all galaxies above our mass resolution 
limit, which is about $10^{8}\msun$, are modeled 
in the entire simulation box, a comparison of the statistical
properties between the SDSS mock catalog and the whole 
simulation box carries information about the CV of the 
SDSS sample.

%%%%%%%%%%%%%%%%%%%%%%%%%%%%%%%%%
%%%%%% part IV. CV of GSMF %%%%%%
%%%%%%%%%%%%%%%%%%%%%%%%%%%%%%%%%
% Yangyao, update, 0807/2018
%   Update figure for corrected GLF/GSMF. 
%   The central CLFs are estimated directly from group catalog.
% Yangyao, update, 0831/2018
%   The corrected GLF/GSMF are updated.

\section{Cosmic variance in galaxy stellar mass functions}
\label{sec_CVinGSMF}

The realistic model catalogs described above have many applications, such 
as to study the relationships between galaxies and the the mass density 
field, and to investigate the galaxy population in different components 
of the cosmic web. Here we use them to analyze and quantify 
the cosmic variances (CV) in the measurements of the galaxy 
stellar mass function (GSMF) and luminosity function (GLF).
We first use model galaxies in the whole simulation 
box to quantify the CV as a function of sample volume and 
galaxy mass. We then use the SDSS mock catalog to examine the 
CV in the SDSS, and to investigate different estimates of the 
GSMF/GLF in their abilities to account for the CV. We propose 
and test a new method that can best correct for the CV. 
Finally, we apply our method to the SDSS catalog to obtain 
GLF and GSMF that are free of the CV. 

\subsection {Cosmic variance as a function of sample volume and galaxy mass}

\begin{figure}
 \includegraphics[width=\columnwidth]{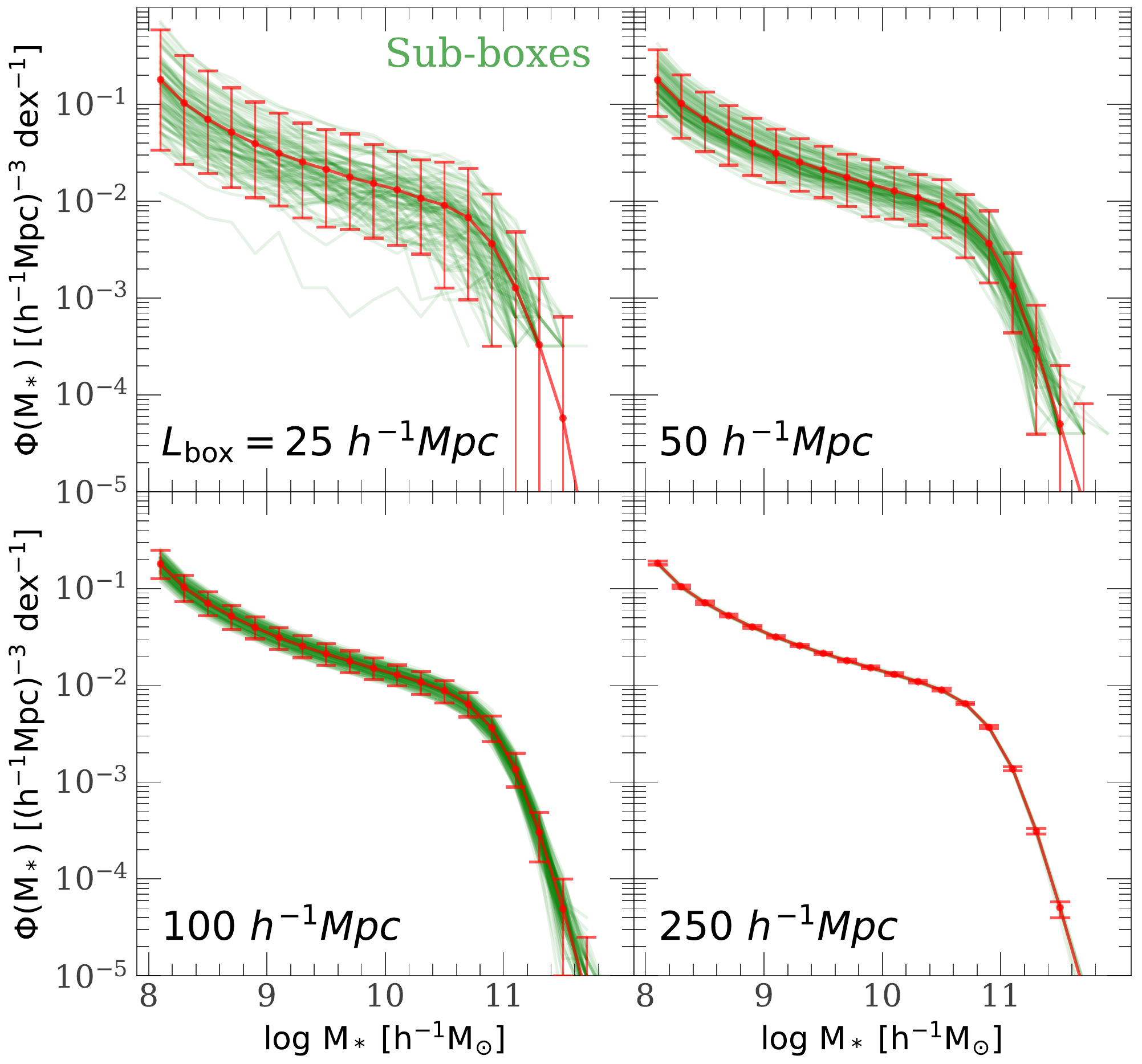}
 \caption{GSMFs at $z=0$ in sub-boxes in the $500 \mpc$ box of ELUCID simulation. 
 For each sub-box size $L_{\rm box}\le 100h^{-1}{\rm Mpc}$, 
 100 sub-boxes without overlap are randomly chosen in the simulation box,
 while for $L_{\rm box} = 250 \mpc$, all the 8 sub-boxes are used.  
 The GSMFs of individual sub-boxes are shown by the green curves 
 in each panel. The average over the sub-boxes of the same  
 size is given by the red line in each panel. Error bars 
 covering $96\%$ ($2\sigma$) range among different sub-boxes 
 are also plotted.}
 \label{fig:cosV_gsmf_boxes}
\end{figure}

\begin{figure}
\centering
 \includegraphics[width=\columnwidth]{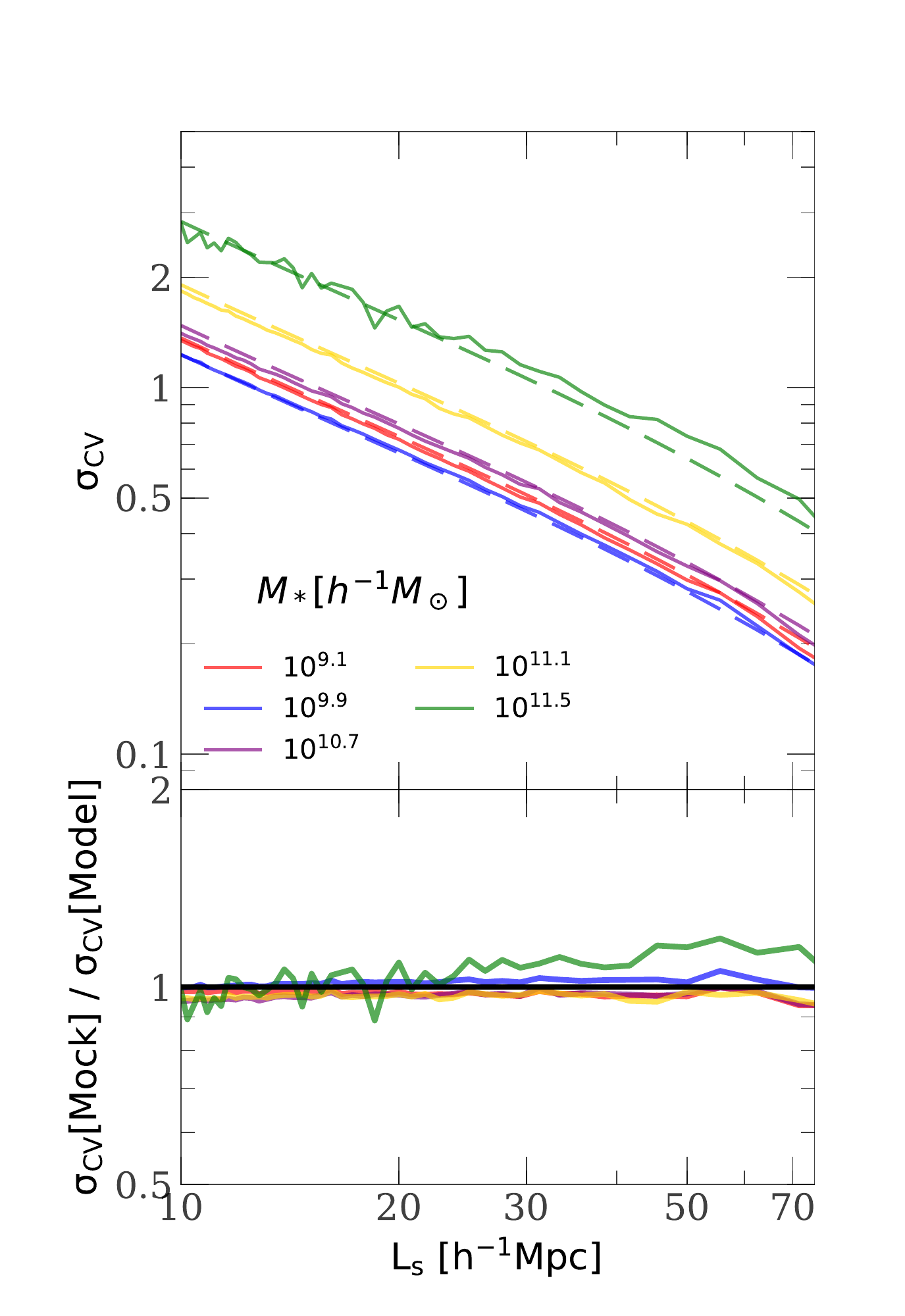}
 \caption{
 Upper panel: Cosmic variance $\sigma_{\rm CV}$ as a function of the 
 characteristic size of sample $L_{\rm s}$ and stellar mass
 $M_*$, as indicated in the panel. Solid lines 
 are $\sigma_{\rm CV}$ estimated from the mock sample, while dashed lines 
 are from the fitting formula. Lower panel: Ratio of $\sigma_{\rm CV}$ between 
 the mock sample and model prediction. 
 The black solid line indicates the ratio of $1.0$.
 }
 \label{fig:cosV_gsmf_box_mock}
\end{figure}

\begin{figure}
 \includegraphics[width=\columnwidth]{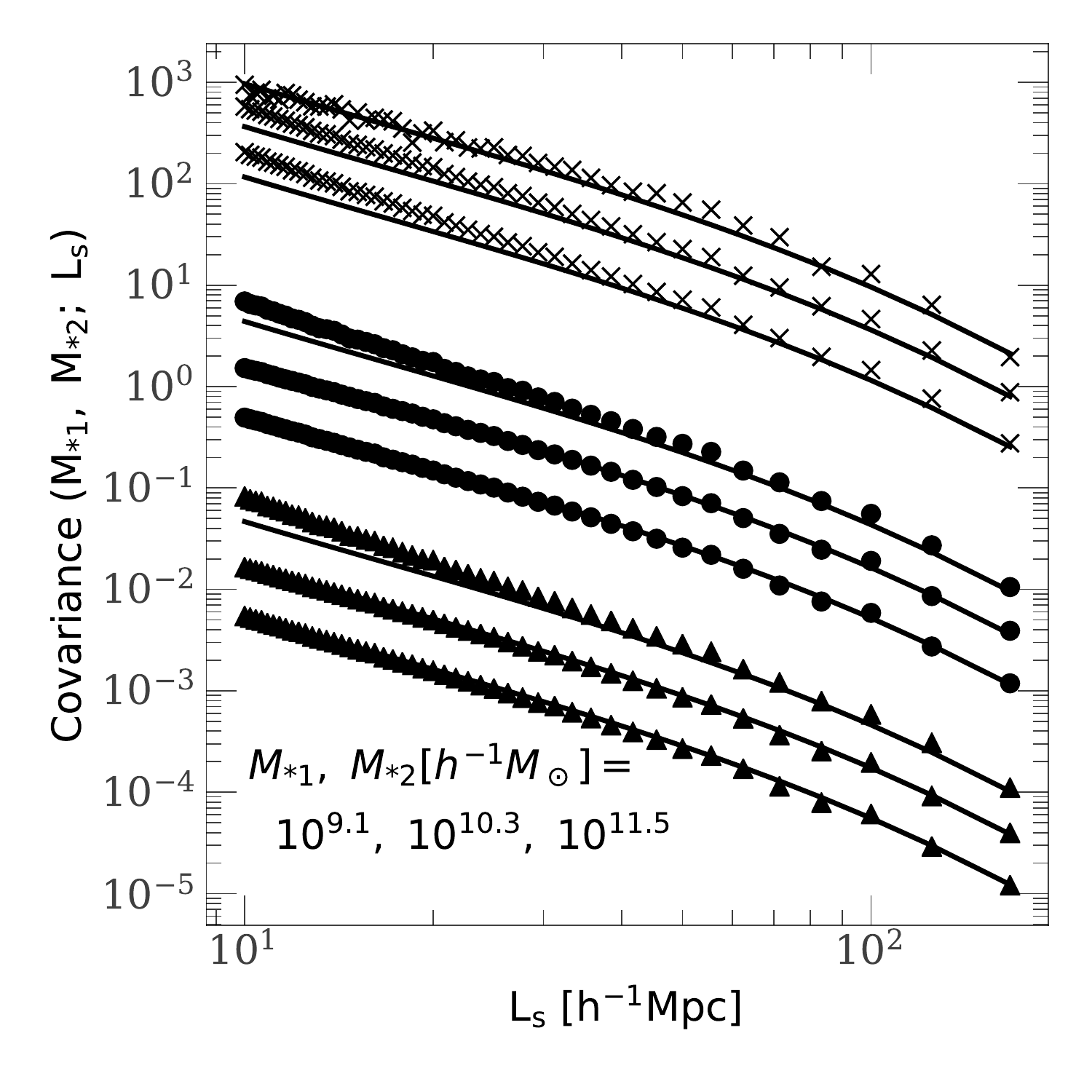}
 \caption{
  The covariance, ${\rm Cov}(M_{*,1},\ M_{*,2};\ L_{\rm s})$,
  of the GSMF between two stellar masses, $M_{*,1}$ and $M_{*,2}$, 
  as a function of the characteristic sample size $L_{s}$.
  Symbols show results obtained from the mock sample. 
  Different $M_{*,1}$ are represented by different symbols:  
  $M_{*,1} [\msun] = 10^{9.1},\ 10^{10.3},\ 10^{11.5}$, from bottom up, 
  scaled by $0.01,\ 1,\ 100$, respectively, for clarity.
  Different $M_{*,2}$ with the same $M_{*,1}$ are re-scaled 
  by $0.3,\ 1,\ 1.2$, for $M_{*,2} [\msun] = 10^{9.1},\ 10^{10.3},\ 10^{11.5}$, 
  respectively. The solid curves are model predictions.
  }
 \label{fig:cov_gsmf_box_mock}
\end{figure}

\begin{figure}
 \includegraphics[width=\columnwidth]{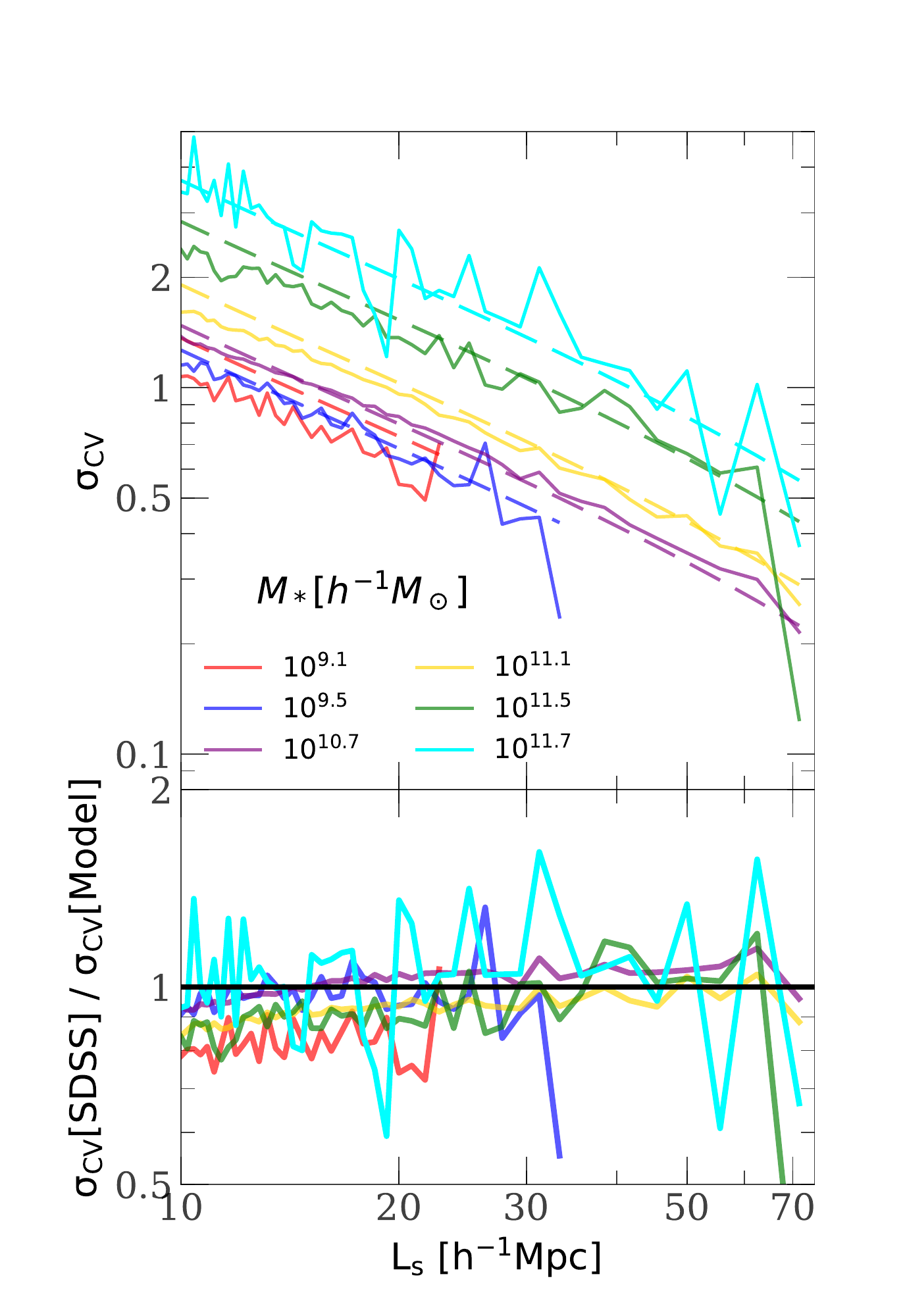}
 \caption{
 The model of cosmic variance compared with SDSS data. 
 Upper panel: Cosmic variance $\sigma_{\rm CV}$ of the GSMF
 as a function of the characteristic sample size $L_{\rm s}$, 
 for galaxies of different stellar masses, $M_*$, as shown by 
 different colors. Solid lines are $\sigma_{\rm CV}$ 
 estimated from the SDSS sample, while dashed lines are predictions 
 of the fitting model.
 Lower panel: The ratio of $\sigma_{\rm CV}$ between the SDSS sample and the 
 model. The black horizontal line indicates the ratio of $1.0$.
 }
 \label{fig:cosV_gsmf_box_SDSS}
\end{figure}

To quantify the effects of CV, we partition the whole 
$500^3\ h^{-3}\mathrm{Mpc}^3$ simulation box 
into sub-boxes, each with a given size $L_{\rm s}$, without overlap. 
For each sub-box $i$ we calculate the galaxy number density 
$n_{g,i}(M_*; L_s)$. Fig.~\ref{fig:cosV_gsmf_boxes} shows the 
GSMFs obtained for 100 sub-boxes with sizes $L_{\rm s} = 25$, $50$, 
$100$, and $250 \mpc$, respectively. The results of individual sub-boxes
are shown by the green lines, while the average and the $2\sigma$ variance 
($96\%$) among the GSMFs are shown by the red curve and bars, 
respectively. As expected, the scatter among the sub-boxes decreases 
as the sub-box size increases. For instance, the scatter for 
$L_{\rm s}=50\mpc$ is about $\approx 0.3\ {\rm dex}$ over almost 
the entire stellar mass range, while it is smaller than $10\%$ 
for $L_{\rm s}=250 \mpc$.  

Theoretically, the galaxy number density $n_g$ is related to 
the mass density $\rho_m$ by a stochastic bias relation:
\begin{equation}
 \delta_g = b \delta_m +\epsilon\,,
\end{equation}
where $\delta_g=(n_g/{\overline n}_g)-1$ and  
$\delta_m=(\rho_m/{\overline\rho}_m)-1$, with 
${\overline n}_g$ and ${\overline\rho}_m$ being the mean 
number density of galaxies and the mean density of mass in the 
Universe. The coefficient, $b$, is the bias parameter, which
characterizes the deterministic part of the bias relation, and 
$\epsilon$ is the stochastic part. If the galaxy number 
density field is a Poisson sampling of the mass 
density field, then the variance in the galaxy density can be 
written as 
\begin{equation}
    \sigma_t^2 = \sigma^2_{\rm CV} + \sigma_{\rm P}^2,
\label{eq_sigma_t}
\end{equation}
where $\sigma_P=N^{-1/2}$ is due to Poisson fluctuation. 
Assuming linear bias, the deterministic part, which we refer
to as the cosmic variance (CV), can be written as:
\begin{equation}
\sigma_{\rm CV}^2 (M_*; L_s) =b^2(M_*) \sigma_m^2 (L_s),
\label{eq_fit_sigma_cv}
\end{equation}
where $L_s$ is the characteristic size of the sample, and 
$\sigma_m(L_s)$ is the rms of the mass fluctuation on the 
scale of $L_s$. 

Motivated by this, we model $\sigma_{\rm CV}$ using the GSMF 
obtained from simulated galaxies. The number density $n_{g,i}$ of 
all sub-boxes are synthesized to give the mean value, 
${\overline n}_g(M_*; L_s)$, and the variance, $\sigma_t^2(M_*;\  L_s)$. 
We use ${\overline n}_g$ to estimate the expected Poisson variance, 
$\sigma_{\rm P}^2$, and use equation (\ref{eq_sigma_t}) to estimate 
$\sigma_{\rm CV}^2(M_*; L_s)$ by subtracting the Poisson part from the
total variance. Equation (\ref{eq_fit_sigma_cv}) is then used 
to fit the dependence of CV on stellar mass and the size of sub-box. 
We find that the $L_s$-dependence can be well described by 
\begin{equation}
\log \sigma_m (x) = p_0 + p_1 x + p_2 x^2 + p_3 x^3\,,  
\end{equation}
where $x = \log (L_{s}/\mpc)$, and 
$p_0=1.53$, $p_1=-2.02$, $p_2=0.92$, and $p_3=-0.25$, 
while the $M_*$ dependence by 
\begin{equation}
\log b(y)= q_0 + q_1 y + q_2 y^2 + q_3 y^3\,,
\end{equation}
where $y = \log (M_*/\msun)$,
and $q_0=-16.04$, $q_1= 5.79$, $q_2=-0.68$, and $q_3=0.026$.

Fig.~\ref{fig:cosV_gsmf_box_mock} shows the comparison between 
$\sigma_{\rm CV}$ obtained direct from the simulated galaxy sample and the 
model prediction as a function of $L_s$ for galaxies of different 
$M_*$, as represented by different lines. The fitting formulae 
work well over the range from $10^8 \msun$ to $10^{11.6} \msun$
in $M_*$, and from $10 \mpc$ to $125\mpc$ in $L_s$.

The above prescription also provides a model for the covariance matrix
of the cosmic variance. Consider the covariance matrix, $C$, 
of the densities between galaxies of masses $M_{*,1}$ and $M_{*,2}$. 
The bias model described above gives 
\begin{equation}
C (M_{*,1}, M_{*,2}; L_s) = b(M_{*,1}) b(M_{*,2}) \sigma_m^2(L_s).    
\end{equation}
Fig.~\ref{fig:cov_gsmf_box_mock} shows the ratio between the measured 
$C$ and the model predictions as a function of $L_s$ for a number of 
$(M_{*,1}, M_{*,2})$ pairs. Overall the model matches the measurements
well. Some discrepancies can be seen for massive galaxies and small 
$L_s$, where the model prediction is slightly lower than that 
measured from the simulation data.

We can compare the CV model calibrated above with that obtained 
from SDSS data. To this end, we estimate the total variance, 
the Poisson variance, and the cosmic variance using sub-boxes of 
given $L_{s}$ that are fully contained by the SDSS 
volume, within which the sample is complete for a given $M_*$. 
In order to estimate the variance among sub-boxes reliably, we
only present cases where at least 10 sub-boxes are available.
The results, plotted in Fig.~\ref{fig:cosV_gsmf_box_SDSS}, 
show that the SDSS measurements follow the model predictions 
for $10 < L_{s} < 75\mpc$ and $ M_* > 10^9\msun $. 
Note that we did not fit the $\sigma_{\rm CV}$ for 
$M_* > 10^{11.6} \msun$, but the extrapolation seems to 
match the SDSS measurements well even for such stellar masses.
For $M_* < 10^9 \msun$, 
the variance obtained from the SDSS becomes significantly 
lower than the model prediction. As we will see below, 
this deviation is caused by the fact that the local 
volume, within which such galaxies can be observed,
does not sample the galaxy population fairly.

To summarize, the simple model presented above provides a useful 
way to estimate the level of CV expected in the 
measurements of the GSMF. This variance, which is produced 
by the fluctuations of the cosmic density field, should be combined with 
the Poisson variance from number counting to estimate the total 
variance in the uncertainty in the GSMF. This is particularly the case 
where the galaxy population is observed in a small volume and the 
cosmic variance is large than the counting error. In real applications, 
other types of uncertainties, such as errors in photometry, 
redshift, and stellar mass estimate, should also be modeled properly
along with the CV described here. 

%%%%%%%%%%%%%%%%%%%%%%%%%%
% CV in SDSS sky coverage
%%%%%%%%%%%%%%%%%%%%%%%%%%
\subsection{Cosmic variances in the SDSS volume}
\label{ssec_cv_sdss_region}

\begin{figure}
\centering
 \includegraphics[width=\columnwidth]{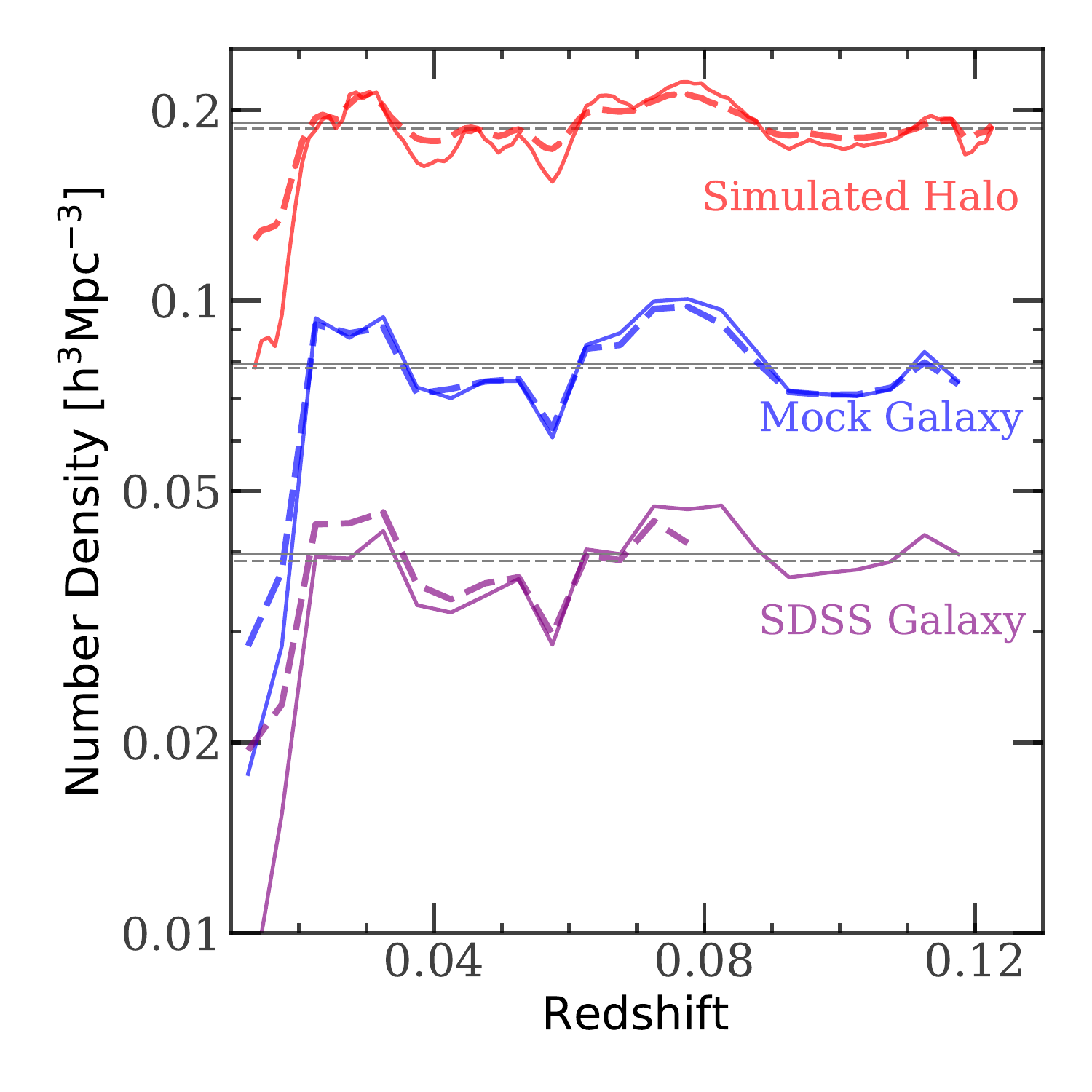}
 \caption{
 Galaxy and halo number densities at different redshift $z$ in the Sloan sky coverage, 
 from $z=0.01$ to $0.12$. Red lines are for simulated halos (Solid: halos with 
 $M_{\rm h}>10^{12}\msun$, offset by $\times 60.0$; Dashed: $10^{8} \leq M_{\rm h} \leq 
 10^{12}\msun$). Blue lines are for mock galaxies based on the empirical model
 (Solid: $M_* > 10^{10.5}\msun$, offset by $\times 22.0$; 
 Dashed: $10^{9.5} \leq M_* \leq 10^{10.5} \msun$, offset by $\times 4.8$). 
 Purple lines are for SDSS galaxies 
 (Solid: galaxies with $M_r < -20.5$, offset by $\times 8.0$; 
 Dashed: $-20.5 \leq M_r \leq -19.5$, offset by $\times 4.4$). 
 Horizontal lines are the mean number densities in the corresponding 
 volumes.
 }
 \label{fig:num_den_4}
\end{figure}

\begin{figure*}
\centering
 \includegraphics[width=14cm]{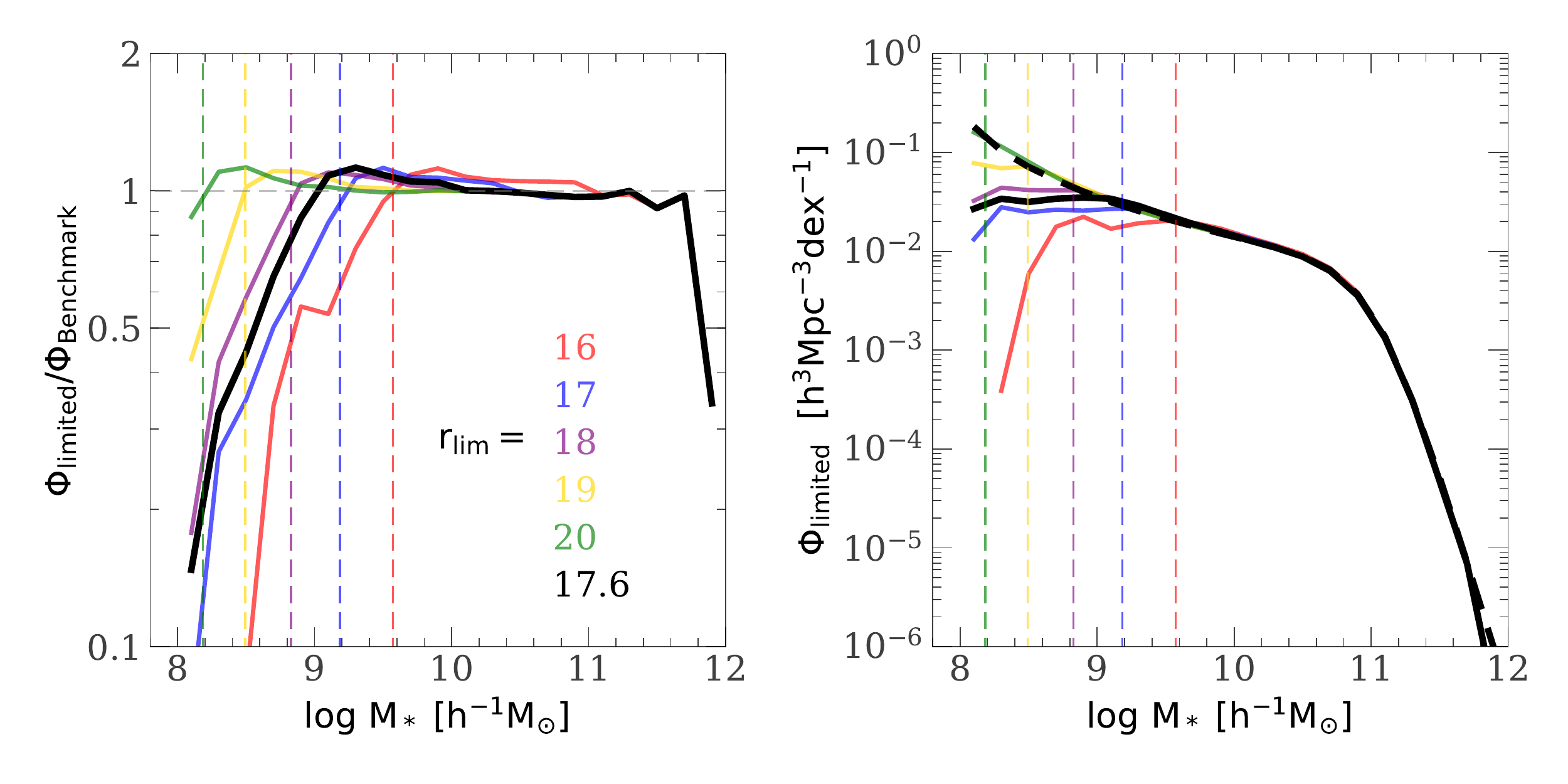}
 \caption{The galaxy stellar mass functions (GSMF), $\Phi(M_*)$, 
 estimated using the V-max method from SDSS magnitude-limited mock samples with different magnitude 
 limits, $r_{\rm lim}$, as shown in the left panel. 
 Right panel shows the absolute values of the GSMF, with  
 the benchmark shown by the black dashed line. The black solid
 shows the result for $r_{\rm lim} =17.6$, the magnitude limit of 
 the SDSS survey. The left panel shows the ratio of GSMF between 
 the magnitude-limited samples and benchmark. 
 In each panel, the vertical dashed lines indicate the stellar masses 
 corresponding to the break at $z = 0.03$.}
 \label{fig:mag_lim}
\end{figure*}

\begin{figure}
\centering
 \includegraphics[width=\columnwidth]{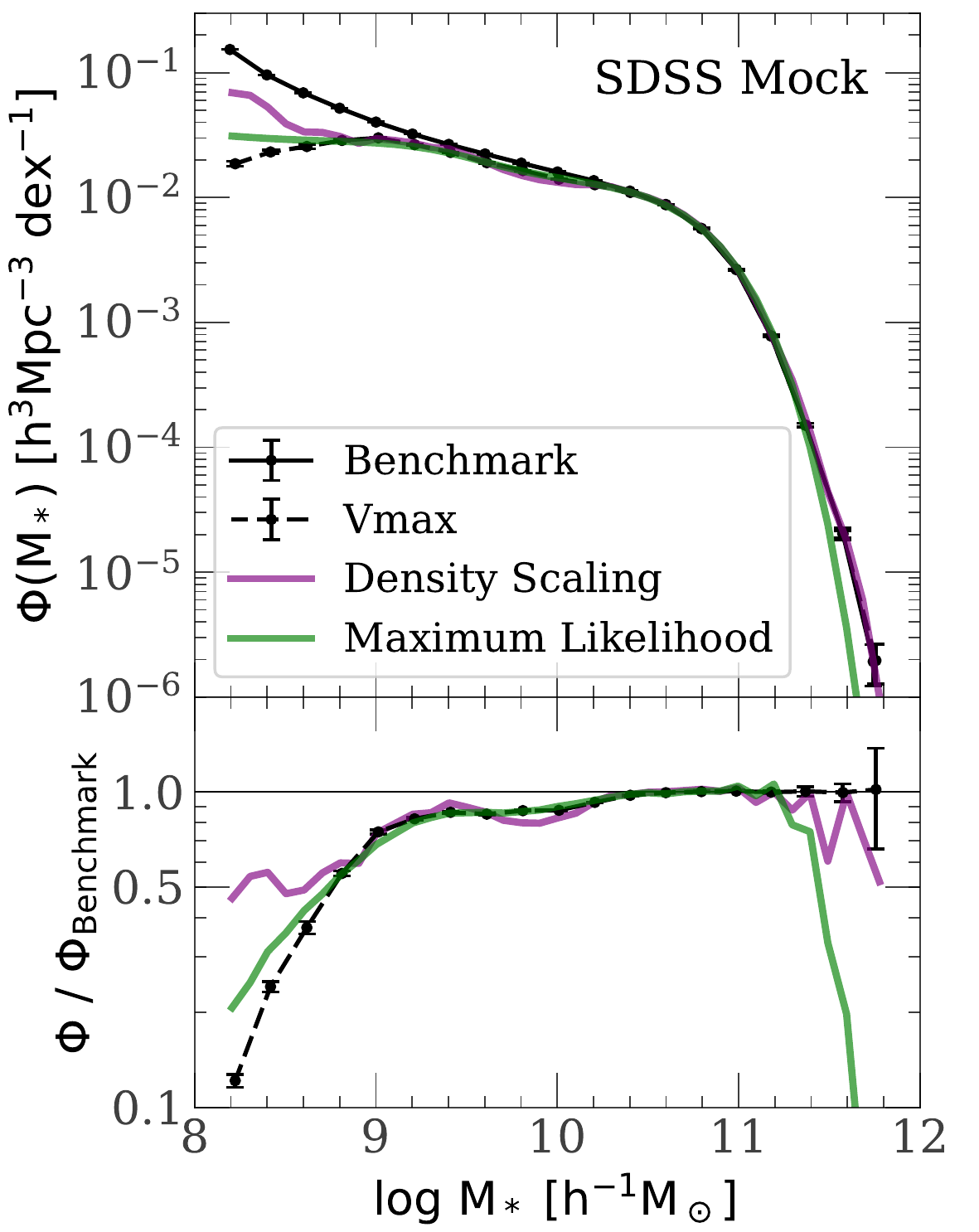}
 \caption{The GSMFs estimated from the SDSS mock catalog
 with different methods designed to account for the cosmic variance. 
 Upper panel shows the GSMFs and lower panel shows the ratio of each GSMF 
 to the benchmark. Black solid line: the benchmark GSMF obtained from the 
 SDSS volume-limited mock sample. Black dashed line: the GSMF obtained with the 
 V-max method, with error bars calculated using 100 bootstrap samples. 
 Purple line: the GSMF obtained from the density scaling method 
 of \citet[][]{Baldry2012}. Green line: the GSMF obtained from the 
 maximum likelihood method assuming a triple-Schechter function form.}
 \label{fig:cor_gsmf_twomethod}
\end{figure} 

In this subsection we examine in detail the CV in the SDSS using 
the mock samples constructed for the SDSS. Here we only consider 
galaxies and model galaxies in the SDSS Northern-Galactic-Cap (NGC) 
(thereafter, SDSS sky coverage) with redshift $0.01 \leq z \leq 0.12$ 
(thereafter, SDSS volume). We construct four different types of samples：
\begin{enumerate}[fullwidth,itemindent=1em,label=(\roman*)]
\item SDSS sample: SDSS DR7 observed galaxies in the SDSS volume, 
      with $r$-band magnitude selection $r \leq 17.6$.
\item SDSS mock sample: model galaxies in the SDSS volume, with r-band 
      magnitude selection $r \leq 17.6$.
\item SDSS magnitude-limited mock samples: model galaxies, in the SDSS volume, 
      that are brighter than a given magnitude limit. 
\item SDSS volume-limited mock sample: all model galaxies in the SDSS volume.     
      This sample is served as the benchmark of the GSMF, since it is almost 
      free of cosmic variance, as compared with the entire simulation box.
\end{enumerate}

To investigate potential cosmic variance in the SDSS volume, we first  
examine the galaxy number density, $n_{g}$, as a function of redshift $z$, 
in the SDSS volume-limited mock sample. Model galaxies in a given stellar 
mass bin are binned in redshift intervals with bin size 
$\delta z = 0.005$, and the galaxy count in each bin is used 
to estimate the galaxy number density. The results are shown in 
Fig.~\ref{fig:num_den_4}. The redshift distribution of model galaxies 
in the SDSS volume shows two peaks, one around  $z_1 \approx 0.03$, 
due to the presence of the large-scale structure known as the CfA Great 
Wall~\citep[][]{Geller1989}, and the other around $z_2 \approx 0.075$
due to the presence of the the Sloan Great Wall \citep[][]{GottIII2005}.
Below $z \approx 0.03$, the number densities show a sharp decline as 
$z$ decreases, and the effect is stronger for massive galaxies,  
indicating the presence of a local void~\citep[see also, for example][]{Whitbourn2014,Whitbourn2016}. 
For comparison, we also show the redshift distribution of 
SDSS galaxies, obtained by using sub-samples complete to given 
absolute magnitude limits. We see that the observed distribution
follows well that in the mock sample, indicating our mock 
sample can be used to study the CV in the SDSS sample. 
For reference we also plot the number densities 
of simulated dark matter halos in the SDSS volume 
versus redshift. Here again we see structures similar to that 
seen in the galaxy distribution. In particular, there is a marked 
decline of halo density at $z< 0.03$, and the decline is more 
prominent for more massive halos.

The presence of the local low-density region shown above can have strong 
impact on the statistical properties of the galaxy population derived 
from the SDSS, especially for faint galaxies which can be 
observed only within the local volume in a magnitude limited sample. Indeed, 
the measurement of the GSMF, which describes the number density of galaxies 
as a function of galaxy mass, can be biased if the 
local low-density region is not properly accounted for. 

As an illustration, Fig.~\ref{fig:mag_lim} shows the GSMFs derived from 
SDSS magnitude-limited mock samples with different $r$-band magnitude 
limits, using the standard $V_{\rm max}$ method. 
For reference, we also plot the GSMF obtained from 
the SDSS volume-limited mock sample (the thick dashed line), 
which matches well the `global' GSMF obtained from the whole 
$500^3\ h^{-3}\mathrm{Mpc}^3$ simulation box. As one can see, the GSMF can be 
significantly underestimated if the magnitude limit is 
shallow (corresponding to a low value of the $r$-band magnitude 
limit, $r_{\rm lim}$). Only a sample as deep as 
$r_{\rm lim}=20$ can provide an unbiased estimate of the 
GSMF down to $M_*\sim 10^8 \msun$. For the SDSS limit, 
$r_{\rm lim}=17.6$, the measurement starts to 
deviate from the global GSMF at $M_* \approx 10^{9} \msun$, and 
the difference between them reaches a factor of about 5
at around $10^8 \msun$. 

 The underestimate of the GSMF at the low-mass end is produced by the 
presence of the low-density region at $z<0.03$ in the SDSS volume. 
To show this more clearly, we define a `break' mass, $M_{0.03}(r_{\rm lim})$, so that 
galaxies with stellar masses $M_* = M_{0.03}$  is complete to 
$z = 0.03$ for the given magnitude limit, $r_{\rm lim}$.
Here we have used the mean mass-to-light ratio, obtained from the mock
sample, to convert the stellar mass to an absolute magnitude. 
As one can see, for each $r_{\rm lim}$, the GSMF obtained 
from the sample starts to deviate from the global benchmark at 
$M_{0.03}(r_{\rm lim})$, shown by the vertical line,  
and is substantially lower at $M_*< M_{0.03}$. All these 
demonstrate that the faint-end of the GSMF can be under-estimated 
significantly in the SDSS due to the presence of the local 
low-density region at $z<0.03$.

\subsection{The correction of cosmic variance}

\subsubsection{Conventional methods}

The results described above indicate that CV is a serious issue 
in the measurements of the GSMF, even for a sample as large as the SDSS. 
Corrections have to be made in order to obtain an unbiased result
that represents the true GSMF in the low-$z$ Universe.  
In the literature, some estimators other than the standard 
V-max method have been proposed, such as the maximum 
likelihood method \citep[e.g.][]{G.Efstathiou1988,Blanton2001,Cole2011,Whitbourn2016}, 
and scaling with bright galaxies \citep[e.g.][]{Baldry2012}.
These methods were designed, at least partly, to correct for the 
effects of large-scale structure in the measurements of the 
GSMF from an observational sample. Here we test their performances 
using our SDSS mock samples. 

 In the maximum likelihood method, one starts with an assumed 
functional form, either parametric or non-parametric, for the 
GSMF, and then use a maximum likelihood method to match the model 
prediction with the data, thereby obtaining the parameters that 
specifies the functional form of the GSMF. In our analysis here,
we choose a triple-Schechter function to model the GSMF,
\beq
        \Phi(M_*) \mathrm{d}\log M_* = \sum_{k=1}^3 \Phi_{*,k}
         \left(\frac{M_*} {\mu_i}\right)^{\alpha_i+1} e^{-M_*/\mu_i}
         {\rm d}\log M_*\,,
\label{eq_triple_schechter}
\eeq
where $\Phi_{*,i}$, $\mu_i$, $\alpha_{i}$ are the amplitude, 
the characteristic mass, and the faint-end slope, of the 
$i$-th Schechter component, respectively. This function is assumed 
to be defined over the domain, $[ M_{*,\rm min},\ M_{*, \rm max} ]$.
For a galaxy, `$i$', with stellar mass $M_i$ at redshift $z_i$ in 
the sample, the probability for it to be observed at this redshift is 
\beq
{\cal L}_i = \frac{\Phi(M_i)}{\int_{M_{i,\rm min}}^{M_{\rm max}}
   \Phi(M_*)\mathrm{d}\log M_*}\,.
\eeq
The total likelihood ${\cal L}$ that the GSMF takes the assumed
$\Phi$ is then given by
\beq
        {\cal L } = \prod_{i=1}^{N} {\cal L}_i\,,
\eeq
where $N$ is the number of galaxies in the sample.
The model parameters can be adjusted so as to maximize the 
likelihood ${\cal L}$. In our application to the SDSS mock sample, 
we fit the GSMF obtained from the V-max method with 
the Triple-Schechter function and use the parameters as 
the initial input of the maximization process. Since the bright end 
is free of cosmic variance, we fix the three parameters 
characterizing the Schechter component at the brightest end, leaving 
the remaining six parameters to be constrained by the maximum 
likelihood process. As the maximum likelihood method does not 
provide information about the overall amplitude of $\Phi(M_*)$, the 
bright end is also used to fix the amplitude of $\Phi(M_*)$.
The GSMF estimated in the way from the SDSS mock sample
is plotted in Fig.~\ref{fig:cor_gsmf_twomethod}
as the green line, in comparison with that estimated by the 
V-max method (dashed line), and the benchmark GSMF (black line).
It is clear that the maximum likelihood method works better 
than the V-max method, but it still underestimates the GSMF 
at the low-mass end. The underlying assumption of the maximum 
likelihood method is that the relative distribution of galaxies 
with respect to $M_*$ is everywhere the same. This in general is 
not true, given that galaxy clustering depends on $M_*$. 
This explains the failure of this method in correcting the CV.

 In an attempt to control the cosmic variance in the GAMA survey, 
\citet{Baldry2012} proposed to use the number density of brighter 
galaxies estimated in a larger volume to scale the number density 
of fainter galaxies that are observed only in a smaller volume.
This method will be referred to as ``density scaling" method. 
Our implementation of this method is as follows. 
\begin{enumerate}[fullwidth,itemindent=1em,label=(\roman*)]
\item 
Choose a `cosmic-variance-free (CVF)' sample, including only bright 
galaxies that have $z_{\rm max}$ larger than $0.12$. In our SDSS mock
sample, this corresponds to select galaxies with 
$M_* >3\times10^{10} \msun$. This sample will be used as the density 
tracer at different redshifts, to scale the density at the fainter end. 
\item 
Compute the cumulative number density of the CVF sample, 
$n_{\rm CVF}(<z)$, as a function of redshift $z$. In practice, 
the cumulative number density is calculated in the 
redshift range $[0.01,\  z]$.
\item
Compute the GSMF, $\Phi_{\rm Vmax}(M_*)$, using the V-max method
\item 
For each stellar mass bin of $\Phi_{\rm Vmax}(M_*)$, find the largest 
redshift, $z_{\rm max}(M_*)$, below which galaxies in this bin can be 
observed in the sample.
\item 
Obtain the corrected GSMF, $\Phi_{\rm sc}$, by scaling the V-max 
estimate with a correction factor: 
\begin{equation}
        \Phi_{\rm sc}(M_*) = \Phi_{\rm Vmax}(M_*) 
        \frac{n_{\rm CVF}(<0.12)}{n_{\rm CVF}[<z_{\rm max}(M_*) ]}\,,     
\end{equation}
where $n_{\rm CVF}(<0.12)$ is the number density of the CVF sample 
in the full redshift range, $[0.01, 0.12]$,  and 
$n_{\rm CVF}[<z_{\rm max}(M_*)]$ is that in the redshift range  
$[0.01,  z_{\rm max}(M_*)]$.
\end{enumerate}

The GSMF estimated in this way from the SDSS mock sample
is plotted in Fig.\,\ref{fig:cor_gsmf_twomethod}
as the purple line. This method appears to work better 
than both the V-max method and the maximum likelihood 
method in the low-mass end, but the underestimation is 
still substantial. Furthermore, this method leads to a dip 
around $M_*=10^{9.8}\msun$, because of the density 
enhancement associated with the CfA Great Wall.    
The failure of this scaling method has an origin similar to 
that of the maximum likelihood method. The underlying 
assumption here is that the bright galaxies can serve as a 
tracer of the cosmic density field, and that the 
distributions of bright and faint galaxies are both related 
to the underlying density field by a similar bias factor. 
In general, this assumption is not valid.

\subsubsection{Methods based on the joint distribution of galaxies 
 and environment}
\label{sssec_mock_test_correction}
 
\begin{figure}
\centering
 \includegraphics[width=\columnwidth]{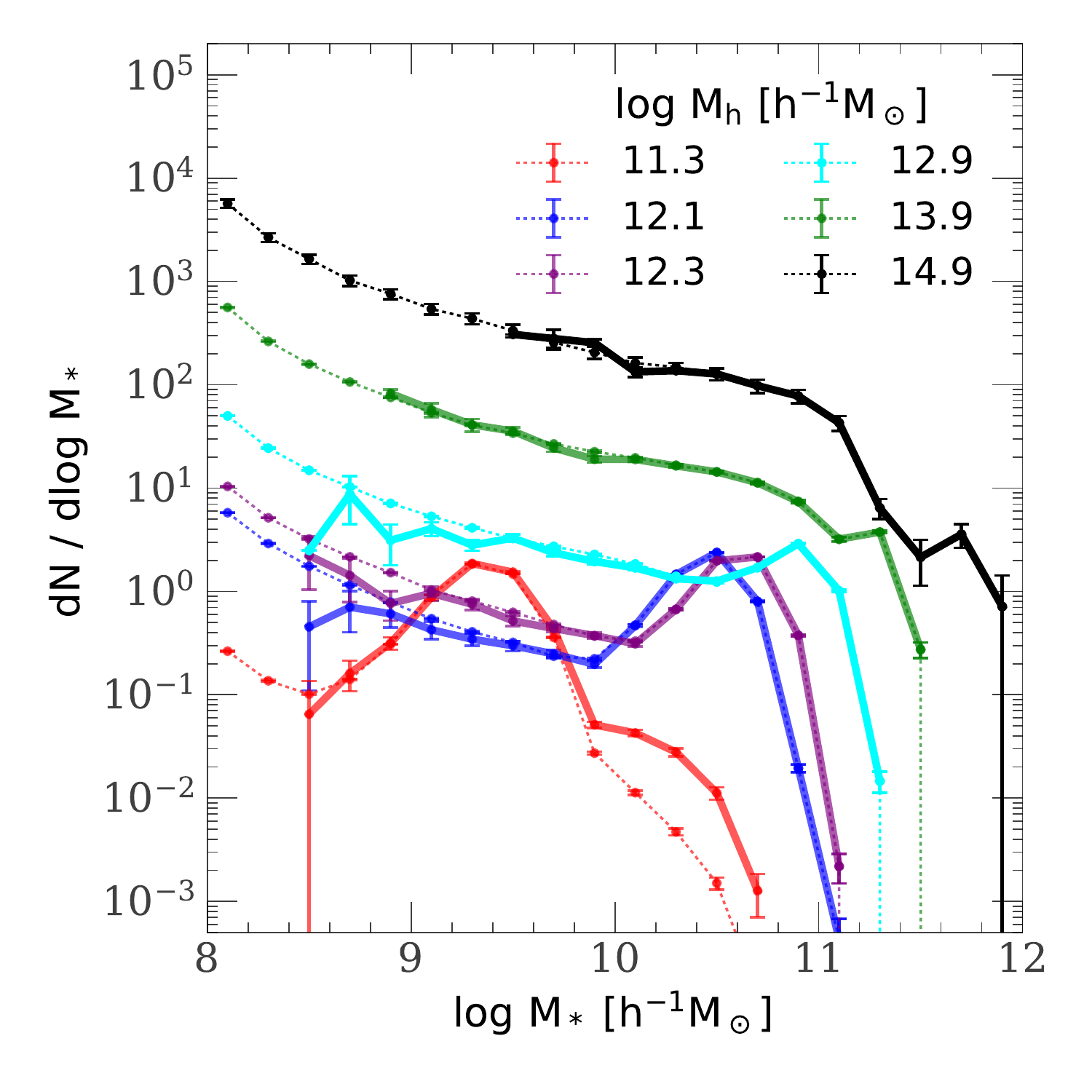}
 \caption{
    The conditional galaxy stellar mass functions (CGSMFs) 
    for halos of different masses, $M_{\rm h}$, as indicated in the figure. 
    Solid lines represent the CGSMFs estimated from the SDSS mock sample. 
    Dotted lines are estimated from the SDSS volume-limited mock sample.
 }
 \label{fig:mockcsmfs}
\end{figure}
 
\begin{figure}
\centering
 \includegraphics[width=\columnwidth]{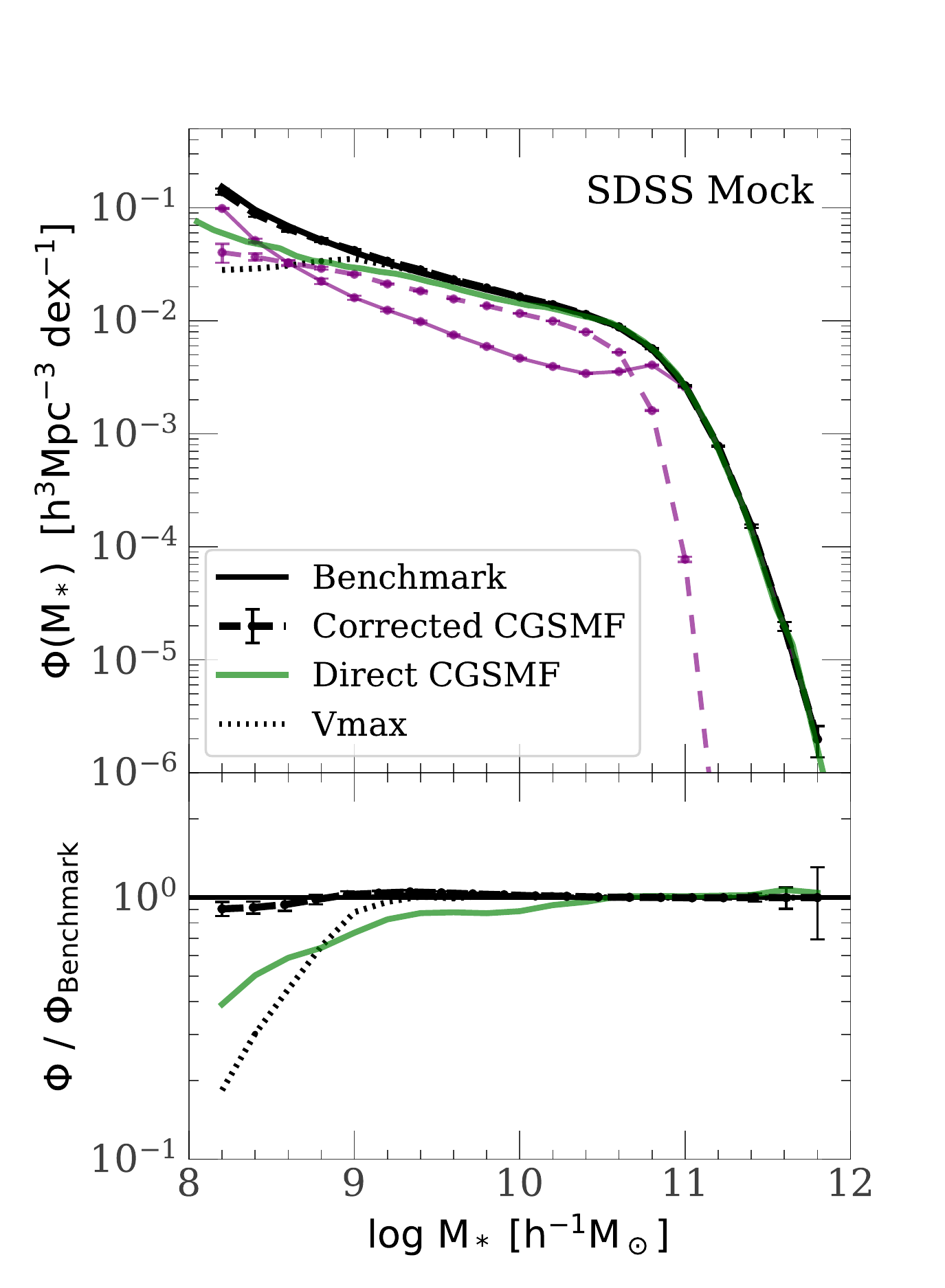}
 \caption{
    The GSMFs obtained by applying different methods to the SDSS mock sample. 
    Upper panel shows the GSMFs while the lower panel shows the ratio of each 
    GSMF to the benchmark.
    Black solid line shows the benchmark GSMF directly calculated from the SDSS
    volume-limited mock sample. Black dashed line is the GSMF obtained by 
    the method based on the CGSMFs described in this paper.
    Black dotted line shows the GSMF derived by using the V-max method. 
    Green line shows the GSMF obtained by combining the CGSMFs directly 
    calculated from the SDSS mock sample (which is incomplete for faint galaxies 
    in massive groups).
    The purple solid and dashed lines are $\Phi_1$ and $\Phi_2$, 
    the contributions of halos with masses $M_{\rm h} \geq 10^{12.5}\msun$ and $M_{\rm h} < 10^{12.5}\msun$, respectively (see~\S\ref{sssec_mock_test_correction} 
    for details). Error bars are calculated from 100 bootstrap samples.
 }
 \label{fig:mock_syn_csmfs}
\end{figure} 

\begin{figure}
\centering
 \includegraphics[width=\columnwidth]{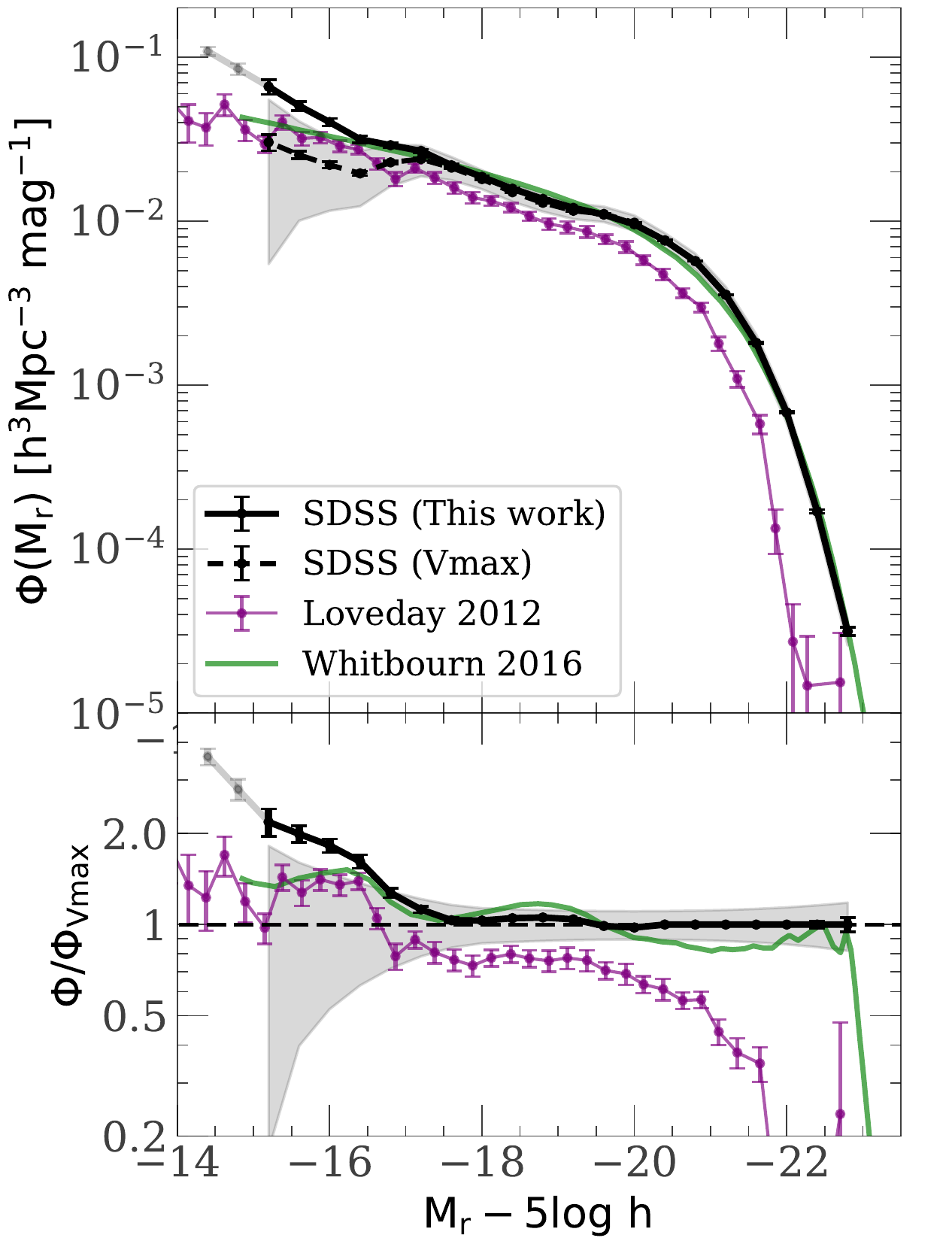}
 \caption{ The galaxy luminosity function (GLF) estimated from the SDSS catalog  
 by using our method, in comparison to the results in the literature.   
 The upper panel shows the GLFs, while the lower panel shows the ratio 
 of each GLF to that obtained with the V-max method.
 The black solid line is the GLF obtained by our method.
 The gray solid line at the 
 faint end (first two data points) are obtained by linear extrapolation. 
 Black dashed line is the GLF by the V-max method. The gray shaded band indicates the cosmic variance of SDSS sample expected from Eq.~\ref{eq_fit_sigma_cv}. 
 Purple line is from~\citet[][]{Loveday2012} for GAMA survey.
 Green line is from~\citet[][]{Whitbourn2016} using SDSS 'cmodel' magnitude.
}
 \label{fig:corrcted_glf}
\end{figure} 

\begin{figure}
\centering
 \includegraphics[width=\columnwidth]{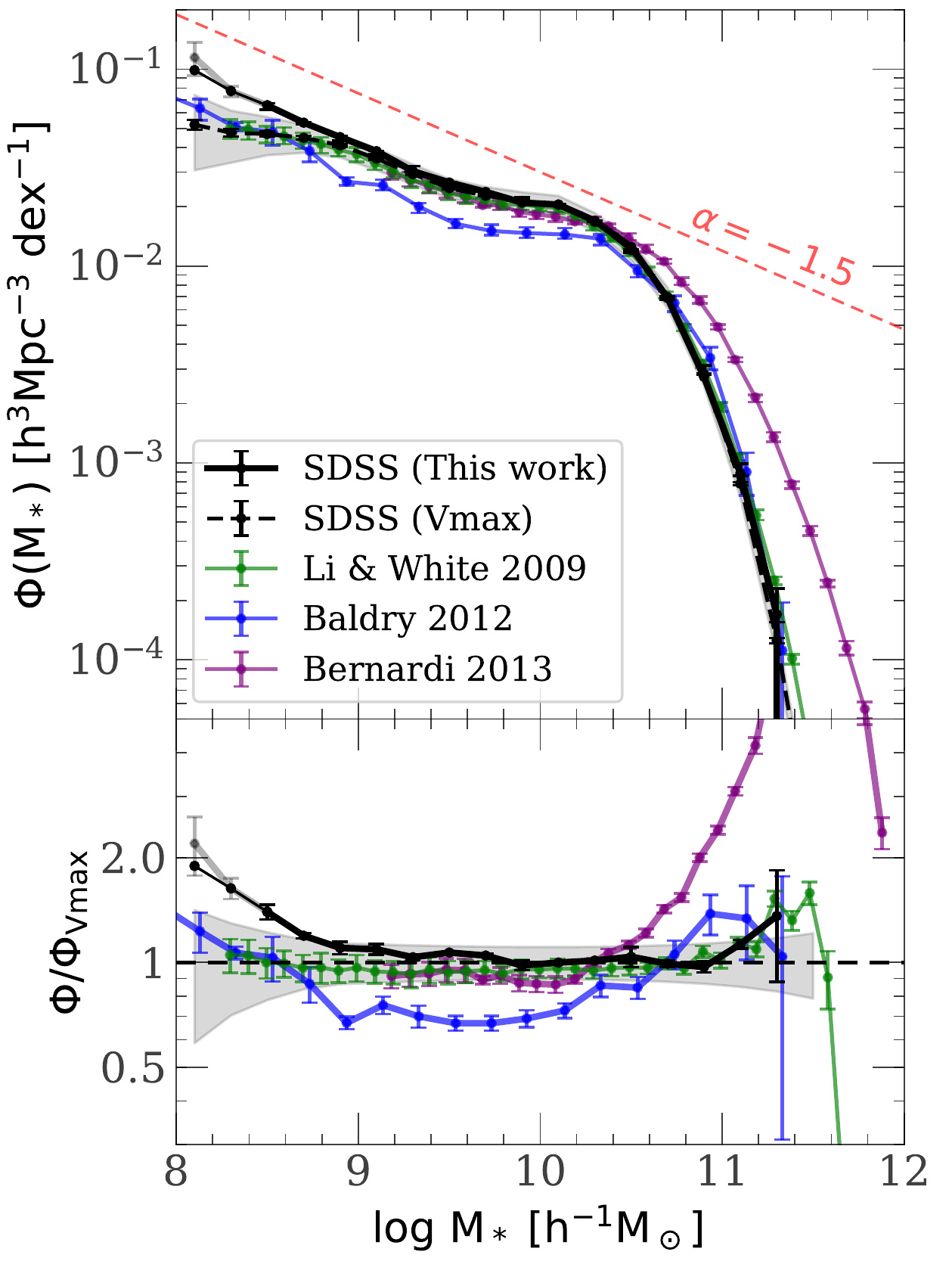}
 \caption{
 The galaxy stellar mass function (GSMF) obtained from the SDSS sample in this paper, in comparison with the results published earlier. 
 Upper panel shows the GSMFs, 
 while lower panel shows the ratio of each GSMF to that given by the V-max method. 
 Black solid line is the GSMF obtained by our method.
 Black dashed line is the GSMF by the V-max method for the SDSS sample.
 The gray shaded band indicates the cosmic variance of SDSS sample expected from Eq.~\ref{eq_fit_sigma_cv}.
 The gray solid piece at the faint end indicates the slight change when 
 the extrapolation of the GLF is used.
 Green line: GSMF from \citet[][]{Li2009}. 
 Blue line: GSMF from the GAMA survey \citep{Baldry2012}. 
 Purple line: GSMF from \citet{Bernardi2013}, 
 where the stellar masses of bright galaxies are estimated from S\'{e}rsic-Exponential fitting.
 The red dashed line in the top of upper panel has a slope $\alpha = -1.5$ (see Eq.~\ref{eq_triple_schechter} for the definition of faint-end slope).
 }
 \label{fig:corrcted_gsmf}
\end{figure}  
 
\begin{table}[b!]
\begin{threeparttable}
\centering
\caption{Corrected galaxy luminosity and stellar mass function}
\label{tab:gsmf}
\begin{tabular}{c c c c c}
\toprule[1.5pt]
$M_{\rm r}-5\log h$ & $\log \Phi(M_{\rm r})$ & $\log M_*$ & \multicolumn{2}{c}{$\log \Phi(M_*)$} \\
 & $[h^3 {\rm Mpc}^{-3} {\rm mag}^{-1}]$ & $[h^{-1}{\rm M_\odot}]$ & \multicolumn{2}{c}{$[h^3 {\rm Mpc}^{-3} {\rm dex}^{-1}]$}\\
\hline
$-15.2$ & $-1.179^{+0.043}_{-0.047}$ & $8.1$ & $-1.005$ & $-0.941$  \\
$-15.6$ & $-1.296^{+0.027}_{-0.028}$ & $8.3$ & $-1.111$ & $-1.111$ \\
$-16.0$ & $-1.395^{+0.022}_{-0.023}$ & $8.5$ & \multicolumn{2}{c}{$-1.184^{+0.011}_{-0.022}$}  \\
$-16.4$ & $-1.500^{+0.022}_{-0.023}$ & $8.7$ & \multicolumn{2}{c}{$-1.272^{+0.003}_{-0.008}$}  \\
$-16.8$ & $-1.536^{+0.015}_{-0.016}$ & $8.9$ & \multicolumn{2}{c}{$-1.345^{+0.007}_{-0.018}$}  \\
$-17.2$ & $-1.571^{+0.011}_{-0.011}$ & $9.1$ & \multicolumn{2}{c}{$-1.417^{+0.001}_{-0.015}$}  \\
$-17.6$ & $-1.659^{+0.009}_{-0.009}$ & $9.3$ & \multicolumn{2}{c}{$-1.518^{+0.026}_{-0.009}$}  \\
$-18.0$ & $-1.730^{+0.006}_{-0.007}$ & $9.5$ & \multicolumn{2}{c}{$-1.576^{+0.002}_{-0.003}$}  \\
$-18.4$ & $-1.802^{+0.006}_{-0.006}$ & $9.7$ & \multicolumn{2}{c}{$-1.624^{+0.000}_{-0.005}$}  \\
$-18.8$ & $-1.864^{+0.005}_{-0.005}$ & $9.9$ & \multicolumn{2}{c}{$-1.678^{+0.026}_{-0.013}$}  \\
$-19.2$ & $-1.919^{+0.004}_{-0.004}$ & $10.1$ & \multicolumn{2}{c}{$-1.688^{+0.004}_{-0.008}$}  \\
$-19.6$ & $-1.960^{+0.003}_{-0.003}$ & $10.3$ & \multicolumn{2}{c}{$-1.770^{+0.000}_{-0.005}$}  \\
$-20.0$ & $-2.019^{+0.002}_{-0.002}$ & $10.5$ & \multicolumn{2}{c}{$-1.903^{+0.022}_{-0.030}$}  \\
$-20.4$ & $-2.117^{+0.002}_{-0.002}$ & $10.7$ & \multicolumn{2}{c}{$-2.163^{+0.010}_{-0.007}$}  \\
$-20.8$ & $-2.244^{+0.002}_{-0.002}$ & $10.9$ & \multicolumn{2}{c}{$-2.562^{+0.058}_{-0.016}$}  \\
$-21.2$ & $-2.449^{+0.003}_{-0.003}$ & $11.1$ & \multicolumn{2}{c}{$-3.053^{+0.049}_{-0.012}$}  \\
$-21.6$ & $-2.743^{+0.004}_{-0.004}$ & $11.3$ & \multicolumn{2}{c}{$-3.770^{+0.131}_{-0.040}$}  \\
$-22.0$ & $-3.166^{+0.006}_{-0.006}$ & & &  \\
$-22.4$ & $-3.770^{+0.011}_{-0.011}$ & & &  \\
$-22.8$ & $-4.501^{+0.023}_{-0.024}$ & & &  \\
\bottomrule[1.5pt]
\end{tabular}
\begin{tablenotes}
\item[$*$] Galaxy brighter than $ M_{\rm r}-5\log h = -15$ is not sufficient to calculate GSMF down to $M_*=10^{8.1},\ 10^{8.3}\msun$. Extrapolation of GLF is used to solve this. Left column of $\Phi(M_*)$ is without extrapolation, while right column is with extrapolation.
\end{tablenotes}
\end{threeparttable}
\end{table}

Since galaxies form and reside in the cosmic density field, 
the number density of galaxies is expected to depend on 
the local environment of galaxies. Suppose the local environment
is specified by a quantity or a set of quantities ${\cal E}$. 
The joint distribution of galaxy mass and ${\cal E}$ 
obtained from a given sample, `$S$', can be written as
\beq
\Phi_S(M_*, {\cal E})
=\Phi_S (M_*\vert {\cal E}) P_S({\cal E}), 
\eeq
where $\Phi_S(M_*\vert {\cal E})$ is the conditional 
distribution of galaxy mass in a given environment estimated from sample 'S', and 
$P_S({\cal E})$ is the probability distribution function of 
the environmental quantity given by the sample. If galaxy formation and 
evolution is a local process so that $\Phi (M_*\vert {\cal E})$
is independent of the galaxy sample, then the CV in the stellar 
mass function derived from the sample can all be attributed 
to the difference between $P_S({\cal E})$ and the global 
distribution function, $P({\cal E})$, expected from a large 
sample where the distribution of ${\cal E}$ is sampled without 
bias. An unbiased estimate of the GSMF $\Phi(M_*)$ is then 
\beq
\label{PhiMstar}
\Phi(M_*)=\int \Phi_S (M_*\vert {\cal E}) P({\cal E}) d{\cal E}\,.
\eeq
Thus, the unbiased GSMF is obtained from the conditional
distribution function, $\Phi_S(M_*\vert {\cal E})$, derived 
from the sample `$S$', and the unbiased distribution $P({\cal E})$ 
of environment variable.

The environmental quantity has to be chosen properly so that 
it can be estimated from observation, while the unbiased distribution 
function, $P({\cal E})$, can, in principle, be obtained from large 
cosmological simulations. Here we analyze a method
which uses the masses of dark matter halos as the environmental 
quantity. In this case, ${\cal E}$ is represented by halo 
mass, $M_{\rm h}$, $\Phi_S (M_*\vert M_{\rm h})$ is the conditional galaxy stellar mass function(CGSMF), and $P({\cal E})=n(M_{\rm h})$ is the halo 
mass function estimated directly from the constrained simulation~\citep{Yang2003}. The advantage here is that the
unbiased estimates are only needed for the conditional functions, 
$\Phi (M_*\vert M_{\rm h})$. The disadvantage is that it is 
model dependent through $n(M_{\rm h})$, and that one has to identify 
galaxy systems to represent dark matter halos. 

Fig.~\ref{fig:mockcsmfs} shows the conditional 
stellar mass functions, of galaxies in halos of different masses, 
estimated from the SDSS mock sample, in comparison with the 
benchmarks obtained from the total SDSS volume-limited sample. 
As one can see, for a given halo mass, the CGSMF obtained 
from the SDSS mock sample matches the benchmark well only 
in the massive end. This happens because of the absence 
of massive halos at small distances in the local under-dense
region, so that their faint member galaxies are not observed in 
the magnitude limited sample. The total GSMF, obtained 
using Eq.~(\ref{PhiMstar}), is shown in the Fig.~\ref{fig:mock_syn_csmfs} 
by the green line, in comparison to the 
benchmark of the total GSMF represented by the black solid line, 
and to the GSMF obtained by the traditional V-max method represented 
by the black dotted line. 
Here the benchmark CGSMFs (dotted lines in Fig.~\ref{fig:mockcsmfs}) 
are used for halos with $M_{\rm h} < 10^{12} \msun$, while the CGSMFs estimated 
from the magnitude-limted sample (solid lines in Fig.~\ref{fig:mockcsmfs}) 
are used for less massive halos. This is to mimic the fact that the total 
CGSMF for less massive halos can be obtained by other means 
(see Eq.~\ref{eq:phi2}), while the low-stellar-mass end of CMGSFs 
for massive halos cannot be obtained directly from the SDSS 
spectroscopic sample.
Here again the stellar mass function 
at the low-mass end is under-estimated, although the method 
works substantially better than the V-max method.
The reason is clear from Fig.~\ref{fig:mockcsmfs}. 
The stellar mass function at the low mass end is only sampled by 
low-mass halos because of the absence of massive halos in the 
nearby volume, while the low-mass end in the benchmark stellar 
mass function is actually affected by the low-mass ends 
of the conditional stellar mass functions of massive halos. 

These results demonstrate an important point. If the shape of the CGSMF depends significantly 
on halo mass, then one needs to estimate all the conditional functions 
reliably down to a given stellar mass limit, in order to
get an unbiased estimate of the total stellar mass function 
down to the same mass limit. The SDSS redshift sample is clearly 
insufficient to achieve this goal in the low-mass end.  

In a recent paper, \citet{Lan2016} showed that the conditional functions of 
galaxies can be estimated down to $M_r\sim -14$ (corresponding to a stellar mass of
about $10^{8}\Msun$) for halos with mass $M_{\rm h}>10^{12}\Msun$ by cross correlating 
galaxy groups (halos) selected from the SDSS spectroscopic sample with SDSS 
photometric data. Thus, if we can estimate the contribution by halos with 
lower masses to a similar magnitude, then the total function can be obtained. 
Here we test the feasibility of such an approach using SDSS mock sample. 
First, we obtain the CGSMFs down to a stellar mass of $10^8\msun$ for halos 
with $M_{\rm h} \geq M_1=10^{12.5}\msun$ directly from the total simulation volume. 
This step is to mimic the fact that such CGSMFs can be obtained, 
as in \citet{Lan2016}, from observational data. 
The GSMF contributed by such halos is
\begin{equation}
\Phi_1 (M_*)
=\int_{M_1}^\infty \Phi (M_*\vert M_{\rm h}) n(M_{\rm h}) d M_{\rm h}\,,
\label{eq:phi1}
\end{equation}
To maximally reduce possible uncertainties introduced by this procedure, 
we estimate the total CGSMF $\Phi_1$ for $M_{\rm h} \geq M_1$ directly 
from a modified V-max method for the high-stellar-mass end.  
Specifically, each galaxy is assigned a weight, 
$n_{\rm halo,u}/n_{\rm halo}(V_{\rm max})$, 
the ratio between the number density of $M_{\rm h}\ge M_1$
halos in the Universe and that in $V_{\rm max}$. 
In practice, the weighted V-max has little impact on the 
results, as the effect of cosmic variance for 
high-mass galaxies is small. The procedure is included 
only for maintaining consistency.
Eq.~(\ref{eq:phi1}) is then used only at the low-stellar-mass end 
where the V-max method fails because of incompleteness. 
The result for $\Phi_1$ obtained in this way is shown by the purple solid 
curve in Fig~\ref{fig:mock_syn_csmfs}. 

To estimate the contribution 
by halos with $M_{\rm h} < M_1$ in a way that can be applied to real observation, 
we first eliminate all galaxies that are contained in halos with 
$M_{\rm h} \geq M_1$. For the rest of the galaxies, we estimate the function by a 
modified version of the V-max method
\begin{equation}
\Phi_2(M_*)
  = \sum { \frac{1}{V_{\rm max}} } \frac{1}{1+b\delta(V_{\rm max})}
\,,
\label{eq:phi2}
\end{equation}
where the summation is over individual galaxies, 
$b = 0.6$ is the bias factor which is considered to be 
constant for low-mass halos \citep[e.g.][]{Sheth2001},  
$\delta(V_{\rm max}) = { {\overline\rho} (V_{\rm max}) / \rho_{\rm u} } - 1$ 
is the mean over density within $V_{\rm max}$, $\rho_{\rm u}$ is the universal 
mass density, and ${\overline\rho}(V_{\rm max})$ is the mean mass density 
within $V_{\rm max}$. The function $\Phi_2$ so estimated  
is shown as the purple dashed curve in Fig.\ref{fig:mock_syn_csmfs}. 
Note that small groups can only be seen in the very local region, 
so the CGSMF estimated for halos in a small mass bin can be very noisy. 
Our method intends to avoid this uncertainty by calculating the total 
CGSMF for all halos less massive than $M_1$. 
The total GSMF, $\Phi=\Phi_1 + \Phi_2$ is shown by the black dashed line in 
Fig.\ref{fig:mock_syn_csmfs}, which is very close to the benchmark, 
indicating that our method can indeed take care of the bias produced by 
the local under-dense region.  We have checked that the result depends 
only weakly on the choice of the value of $M_1$. 

\subsection{Applications to observational data}

In this subsection, we apply the method described above to the real SDSS sample. 
We first estimate the galaxy luminosity function (GLF) using 
the procedure based on the conditional distributions of galaxy luminosity in 
dark matter halos, as described in \S\ref{sssec_mock_test_correction}. 
Here the CLF, $\Phi_1(M_r)$, for faint galaxies with magnitude 
$M_r-5\log h > -17.2$ in halos more massive than $10^{12.5}\msun$ are 
obtained from \citet{Lan2016}, while the CLF for brighter galaxies 
in these halos is estimated  directly from SDSS sample using the 
V-max method and the group catalog of \citet[][]{Yang2012}~\citep[see also][]{Yang2007}. 
For halos with masses below $10^{12.5}\msun$ the CLF, $\Phi_2(M_r)$, 
is obtained from the SDSS sample using the modified V-max method as 
described by Eq.\,(\ref{eq:phi2}). The total galaxy luminosity 
function (GLF) is then obtained by $\Phi(M_r)=\Phi_1(M_r)+\Phi_2(M_r)$. 
Fig.~\ref{fig:corrcted_glf} shows the result of the GLF so obtained 
in solid black line, in comparison with that obtained from the traditional V-max method. 
At the faint end, $M_r-5\log h \approx -15$, the GLF is about twice 
as high as that given by the V-max method, indicating that 
cosmic variance can have large impact on the estimate of the 
GLF at the faint end.
To show this more clearly, we plot the cosmic variance expected from 
Eq.~\ref{eq_fit_sigma_cv} as the shaded band in Fig.~\ref{fig:corrcted_glf}, 
where the stellar mass is obtained from luminosity by 
using mean mass-to-light ratio. The expected cosmic variance is 
quite large at the faint end, indicating that cosmic variance is an 
important issue in estimating the faint end of GLF.
The GLF in the local Universe has been estimated by many
authors using various samples 
\citep[e.g.][]{Blanton2003,Yang2009,Loveday2012,Jones2006,Driver2012,Whitbourn2016}. 
For comparison, we plot the GLFs obtained by \citet{Loveday2012}
from the GAMA survey and by \citet{Whitbourn2016}, who applies a maximum 
likelihood method to the SDSS to account for the cosmic variance in their 
estimates. The result of \citet{Whitbourn2016} matches ours 
over a wide range of luminosity, but seems to still  
underestimate the GLF at the faint end.
The result of \citet{Loveday2012} has a large discrepancy with our 
result, possibly due to the cosmic variance in the small sky 
converage of the GAMA sample used, which is $144\deg^2$. 
Since many of the faint galaxies in the SDSS photometric data 
do not have reliable stellar mass estimates, conditional 
galaxy stellar mass functions are not available at the low 
mass end. Because of this, we cannot estimate the GSMF 
down to the low-mass end directly from the data with the 
method above. As an alternative, we use the $M_r$ - $M_*$
relation obtained from the SDSS spectroscopic sample 
to convert the GLF obtained above to estimate a GSMF.  
We do this through the following steps. (i) Construct a large 
volume-limited Monte-Carlo sample of galaxies with absolute magnitude 
distribution given by the GLF. (ii) Bin these galaxies according 
to their absolute magnitudes. (iii) For each Monte-Carlo galaxy, 
we randomly choose a galaxy in the real SDSS spectroscopic sample 
in the same absolute magnitude bin, and assign the stellar 
mass of the real galaxy to the Monte Carlo galaxy. 
(iv) Compute the GSMF of this volume-limited Monte-Carlo sample. 

The GSMF obtained directly from the GLF in this way 
is shown in Fig.~\ref{fig:corrcted_gsmf} by the black solid curve.
Since the GLF is estimated only down to $M_r-5\log h \approx -15$, 
the first two data points in the low-mass end of the GSMF
may be underestimated, as galaxies fainter than 
$M_r-5\log h = -15$ may contribute to these two stellar mass bins.
To test this, we extrapolate the faint end of 
the GLF to $M_r-5\log h = -14.2$, which is sufficient 
to include all galaxies with stellar masses down to 
$10^8\msun$. This extrapolation is shown by the gray extension of 
the black solid curve in Fig.~\ref{fig:corrcted_glf}.
The GSMF obtained from the extended GLF is shown by the gray line in~Fig.\ref{fig:corrcted_gsmf}.  
As one can see, the extension of the GLF only slightly increases the GSMF 
at the lowest mass. The GSMF estimated in this way is compared 
with that estimated with the conventional V-max method. 
The gray shaded band shows the expected cosmic variance 
given by Eq.~\ref{eq_fit_sigma_cv} for the SDSS sample.
The effect of CV is quite large at the low-stellar-mass end. 
The difference between our result and that obtained 
from the V-max method is even larger, indicating again 
that the local SDSS region is an unusually under-dense region.
The GSMF in the low-$z$ Universe has been estimated in 
numerous earlier investigations using different samples and 
methods \citep[e.g.][]{Li2009,Yang2009,Baldry2012,Bernardi2013,He2013,DSouza2015}. 
Several of the earlier results are plotted in 
Fig.~\ref{fig:corrcted_gsmf} for comparison. The result of 
\citet{Li2009}, who measured the GSMF of SDSS sample directly from the 
stellar masses estimated by \citet{Blanton2007}
with a Chabrier IMF~\citep[][]{Chabrier2003} and corrections
for dust, matches well our V-max result, and also misses 
the steepening of the GSMF at $M_*< 10^{9.5} \msun$. 
Our measurement at $M_*>10^{10.5}\msun$ is significantly lower 
than that from \citet{Bernardi2013}, because they included  
the light in the outer parts of massive galaxies that may be missed 
in the SDSS NYU-VAGC used here (see also \citealt{He2013} 
for the discussion of this effect). Such corrections do not affect the GSMF 
in the low-mass end. The overall shape of our GSMF is similar to that of 
\citet{Baldry2012} obtained from the GAMA sample, 
but the amplitude of their function is about $50\%$ lower. 
GAMA has a small sky coverage, $144\deg^2$, although it is deeper, 
to $r\approx 19.8$. According to our test with mock samples of a 
similar sky coverage and depth, the cosmic variance in 
the GSMF estimated from such a sample can be very large. 
The lower amplitude given by the GAMA sample may be produced by 
such cosmic variance.

In conclusion, when CV is carefully taken into account, 
the low-stellar-mass end slope of the GSMF in the low-$z$ Universe, 
which is about $-1.5$ as indicated by the red dashed line 
in Fig.~\ref{fig:corrcted_gsmf}, is significantly 
steeper than those published in earlier studies. In particular, 
there is a significant upturn at $M_*< 10^{9.5} \msun$ 
in the GSMF that is missed in many of the earlier measurements.
For reference, we list the GLF and GSMF estimated with our method 
in Table~\ref{tab:gsmf}.

%%%%%%%%%%%%%%%%%%%%%%%%%%%%%%%%%%%%%%
%%%%%%   part VI. summary %%%%%%%%%%%%
%%%%%%%%%%%%%%%%%%%%%%%%%%%%%%%%%%%%%%
% Yangyao, update, 0807/2018

\section{Summary and discussion}
\label{sec_summary}

 In this paper, we use ELUCID simulation, a constrained $N$-body 
simulation in the Sloan Digital Sky Survey (SDSS) volume to 
study galaxy distribution in the low-$z$ Universe. Our main results 
can be summarized as follows:
\begin{enumerate}[fullwidth,itemindent=1em,label=(\roman*)]
\item 
Dark matter halos are selected from different snapshots 
of the simulation, and halo merger trees are constructed 
from the simulated halos down to a halo mass of $\sim 10^{10}\msun$.
A method is developed to extend all the simulated halo merger 
trees to a mass resolution of $10^9\msun$, which is needed to 
model galaxies down to a stellar mass of $10^8\msun$.  
\item
The merger trees are used to populate simulated dark matter halos
with galaxies according to an empirical model of galaxy formation
developed by \citet{Lu2014a,Lu2015a}. The model galaxies 
follow the real galaxies in the SDSS volume both in spatial 
distribution and in intrinsic properties. The catalog of the  
model galaxies, therefore, provide a unique way to study galaxy 
formation and evolution in the cosmic web in the low-$z$ Universe.
\item
Mock catalogs in the SDSS sky coverage are constructed, which can be 
used to investigate the distribution of galaxies as measured
from the real SDSS data and its relation to the global distribution 
expected from a fair sample of galaxies in the low-$z$ Universe. 
These mock catalogs can thus be used to quantify the cosmic 
variances in the statistical properties of the low-$z$ galaxy 
population estimated from a survey like SDSS. 
\item 
As an example, we use the mock catalogs so constructed to quantify
the cosmic variance in the galaxy stellar mass function (GSMF). 
Useful fitting formulae are obtained to describe the cosmic variance and covariance 
matrix of the GSMF as functions of stellar mass and 
sample volume. 
\item
We find that the GSMF estimated from the SDSS 
magnitude-limited sample can be affected significantly by the 
presence of the under-dense region at $z<0.03$, so that the low-mass 
end of the function can be underestimated significantly. 
\item 
We test several existing methods that are designed to deal with the 
effects of the cosmic variance in the estimate of GSMF, and 
find that none of them is able to fully account for the cosmic 
variance effects. 
\item
We propose and test a method based on the conditional stellar mass functions  
in dark matter halos, which is found to provide an unbiased estimate of the 
global GSMF.   
\item 
We apply the method to the SDSS data and find that the 
GSMF has a significant upturn at $M_*< 10^{9.5} \msun$, which is missed 
in many earlier measurements of the local GSMF.
\end{enumerate}

Our results of the GSMF have important implications for galaxy formation
and evolution. The presence of an upturn in the GSMF at $M_*<10^{9.5}\msun$
suggests that there is a characteristic mass scale, $\sim 10^{9.5}\msun$, 
corresponding to a halo mass of $\sim 10^{11}\msun$ \citep[e.g.][]{Lim2017a},
%Lim, S. H.; Mo, H. J.; Lan, T.-W.; Ménard, B.  2017MNRAS.464.3256L
below which star formation may be affected by processes that are 
different from those in galaxies of higher masses. The stellar mass 
function of galaxies at low-$z$ has been widely used to calibrate 
numerical simulations and semi-analytic models of galaxy formation. 
The improved estimate of the GSMF presented here clearly will provide 
more accurate constraints on theoretical models. 

The mock catalogs constructed here have other applications. 
For example, they can be used to analyze the cosmic variance in the 
measurements of other statistical properties of the galaxy population, 
such as the correlation functions \citep{Zehavi2005, WangYu2007}
and peculiar velocities \citep[e.g.][]{Jing1998,Loveday2018}
of galaxies of different luminosities/masses. Because of the presence 
of local large-scale structures, such as the under-dense region 
at $z<0.03$, the measurements for faint galaxies can be affected.
A comparison between the results obtained from the mock sample 
and that from the benchmark sample can then be used to 
quantify the effects of cosmic variance. Another application 
is to HI samples of galaxies. Current HI surveys, 
such as HIPASS \citep{Meyer2004} and ALFALFA \citep{Giovanelli2005}, 
are shallow, typically to $z\sim 0.05$, and so 
the HI-mass functions and correlation functions estimated 
from these surveys can be affected significantly by the cosmic 
variance in the nearby Universe \citep[e.g.][]{Guo2017}. 
The same method as described here can be used to construct 
mock catalogs for HI galaxies, and to quantify cosmic 
variances in these measurements. We will come back to some of 
these problems in the future.

\section*{Acknowledgements}
This work is supported by the National Key R\&D Program of China (grant Nos. 2018YFA0404503, 2018YFA0404502), the National Key Basic Research Program of China (grant Nos. 2015CB857002, 2015CB857004), and the National Science Foundation of China (grant Nos. 11233005, 11621303, 11522324, 11421303, 11503065, 11673015, 11733004, 11320101002). HJM acknowledges the support from NSF AST-1517528.

\bibliography{MyCollection.bib}
\end{document}